\documentclass{article}

\usepackage{PRIMEarxiv}

\usepackage[utf8]{inputenc} 
\usepackage[T1]{fontenc}    
\usepackage{hyperref}       
\usepackage{url}            
\usepackage{booktabs}       
\usepackage{amsfonts}       
\usepackage{nicefrac}       
\usepackage{microtype}      
\usepackage{lipsum}
\usepackage{fancyhdr}       
\usepackage{graphicx}       
\graphicspath{{media/}}     

\title{Bayesian Covariance Structure Modeling of Multi-Way Nested Data
}

\author{
  Stef Baas, Richard J. Boucherie, Jean-Paul Fox \\
  University of Twente \\
  Enschede}

\usepackage{amsthm}
\usepackage{amsmath}

\usepackage{bm}
\usepackage{amsfonts}
\usepackage{graphicx}

\newtheorem{theorem}{Theorem}
\newtheorem{lemma}[theorem]{Lemma}
\newtheorem{corollary}[theorem]{Corollary}
\newtheorem{Remark}{Remark}

\def\mydefb#1{\expandafter\def\csname b#1\endcsname{\mathbf{#1}}}
\def\mydefallb#1{\ifx#1\mydefallb\else\mydefb#1\expandafter\mydefallb\fi}
\mydefallb ABCDEFGHIJKLMNOPQRSTUVWXYZeucsxbrvht\mydefallb

\def\mydefgreek#1{\expandafter\def\csname b#1\endcsname{\text{\boldmath$\mathbf{\csname #1\endcsname}$}}}
\def\mydefallgreek#1{\ifx\mydefallgreek#1\else\mydefgreek{#1}%
	\lowercase{\mydefgreek{#1}}\expandafter\mydefallgreek\fi}
\mydefallgreek {beta}{Gamma}{Delta}{ell}{epsilon}{Omega}{etaex}{Theta}{Iota}{Lambda}{kappa}{mu}{nu}{Xi}{Pi}{rho}{Sigma}{tau}\mydefallgreek

\usepackage{algorithm,algpseudocode}
\algnewcommand{\Inputs}[1]{%
	\State \textbf{Inputs:}
	\Statex \hspace*{\algorithmicindent}\parbox[t]{.8\linewidth}{\raggedright #1}
}
\algnewcommand{\Initialize}[1]{%
	\State \textbf{Initialize:}
	\Statex \hspace*{\algorithmicindent}\parbox[t]{.8\linewidth}{\raggedright #1}
}
\usepackage{setspace}
\usepackage{placeins}
\newcommand{\ubar}[1]{\text{\b{$#1$}}}
\begin{document}

\maketitle

\begin{abstract}
A Bayesian multivariate model with a structured covariance matrix for multi-way nested data is proposed. This flexible modeling framework allows for positive and for negative associations among clustered observations, and generalizes the well-known dependence structure implied by random effects. A conjugate shifted-inverse gamma prior is proposed for the covariance parameters which ensures that the covariance matrix remains positive definite under posterior analysis. A numerically efficient Gibbs sampling procedure is defined for balanced nested designs, and is validated using two simulation studies. For a top-layer unbalanced nested design, the procedure requires an additional data augmentation step. The proposed data augmentation procedure facilitates sampling latent variables from (truncated) univariate normal distributions, and avoids numerical computation of the inverse of the structured covariance matrix. The Bayesian multivariate (linear transformation) model is applied to two-way nested interval-censored event times to analyze differences in adverse events between three groups of patients, who were randomly allocated to treatment with different stents (BIO-RESORT). The parameters of the structured covariance matrix represent unobserved heterogeneity in treatment effects and are examined to detect differential treatment effects. 
\end{abstract}

\keywords{Bayesian probit model \and Covariance structure model \and interval-censored times \and multi-way nested design \and shifted-inverse gamma distribution}


\section{ Introduction}
	Multilevel or hierarchical structured outcomes occur frequently in various research disciplines. For instance, these structures can be found in studies of the development of physical symptoms, genetic disease among family members, studies with multi-arm multi-stage designs, multi-centre trials and in experiments with multiple endpoint measurements. Measurements of patients can be considered to be clustered, where clustered observations are correlated since they share some homogeneous features (e.g., they are obtained from the same patient or from patients receiving the same treatment). The analysis of multi-way nested measurements can be complicated, since they are often measured on a discrete scale (e.g., event time, binary, ordered categorical). Before discussing the merits of our Bayesian multivariate modeling approach for multi-way nested data, the deficiencies of multilevel approaches are discussed.  

	\subsection*{Failure of multilevel approaches}   
	Traditionally, the modeling of multi-way nested data is done using latent variables, also referred to as random effects, where the dependence among clustered observations is modeled by the sharing of a random effect. This approach has been popularized in many ways, for instance, through factor analysis models, multilevel models, and frailty survival models. However, this popular approach has several disadvantages. 
	\newpage
	First, random effects can only describe positive within-cluster correlation, since it is based on shared/pooled information. The pooling of information across clusters is operated through random effects, and clustered observations are positively correlated by the sharing of a random effect. On top of that, there are only a few statistical approaches to model negative within-cluster correlation, which are limited to small clusters         \cite{kenny2002statistical,nielsen2021small,snijders1999social}. 
	
	Second, the random effect variance has a natural lower-bound of zero. When the random effect variance approaches this lower bound, estimation methods behave poorly and often fail to converge, since the covariance matrix of the random effects becomes singular \cite{gelman2008using,gilks1996strategies,Harville1977,zeger1991generalized}.	The likelihood function becomes intractable when the covariance matrix is singular. Because a random effect variance of zero  represents a singularity, a spike-and-slab prior has been proposed to include this point in the parameter space under the model \cite{FrhwirthSchnatter2010BayesianVS}. One component represents zero variance (the spike)  and the second component (the slab) represents non-zero random effect variance. However, this two-component mixture prior makes Bayesian inference more difficult, it increases the computational burden, and its performance is sensitive to hyper-parameter settings.         
	
	Third, the significance of a random effect variance is often of specific interest. However, this point is on the boundary of the parameter space, and commonly used tests, such as the likelihood ratio, Wald and score tests, do not have the traditional chi-squared distribution \cite{self1987asymptotic,stram1994variance,feng1992statistical}. Furthermore, the sampling distribution of variance estimates is strongly asymmetric, which makes the standard error a poor characterization of the uncertainty. When testing the significance of a random effect variance with a likelihood ratio test, the P-value is conservative and approximately twice as large as it should be \cite{pinheiro2006mixed}. Currently, there is no widely accepted (parametric) test procedure to test the significance of a random effect variance \cite{rao2019permutation}. Developed methods are computationally intensive and provide approximate results \cite{pauler1999bayes,saville2009testing}. Permutation tests have been proposed as a nonparametric method to test variance components \cite{fitzmaurice2007note}, but they may not be transferable to more complex nested data structures with crossed random effects and are also computationally intensive.
	
	Fourth, the number of model parameters increases with the number of clusters. Therefore, the complexity of the model increases when collecting more data, which makes the modeling approach not suitable for high-dimensional large data \cite{perry2017fast}. 
	
	Fifth, the computational burden is high for multi-way nested discrete data, since it requires integrating over a high-dimensional truncated parameter space. For instance, the multivariate analysis of discrete response data through multiple latent variables, representing a confirmatory factor analysis model, has become very popular and different estimation methods have been proposed (see e.g., \cite{muthen1979structural,bock1996high,song2005multivariate}). Although models of this type are very flexible, parameter estimation can be highly computationally intensive, when more than a few latent variables are included \cite{cai2010high,schilling2005high,wirth2007item}. 
	
	\subsection*{Bayesian modeling of covariance structures}
	In order to overcome these disadvantages, an integrated likelihood approach is followed \cite{berger1999integrated}. The random effects are considered nuisance parameters and are integrated out. As a result, the dependence structure, which is implied by the random effects, is directly modeled with a multivariate linear model and a structured covariance matrix. Our Bayesian multivariate model with a structured covariance matrix is referred to as a Bayesian covariance structure model (BCSM), in which associations among multi-way nested data are efficiently represented by covariance parameters. 
    
    Early work on covariance structure analysis considered a multivariate normal distribution for the observed data with a parametric form for the covariance matrix \cite{bock1966analysis,joreskog1970general,wiley1973studies}. The parametric form for the covariance matrix represents the dependence structure of a linear factor analytic model with normally distributed random effects (referred to as latent factors or latent variables). The objective was to obtain maximum likelihood estimates for the variance components associated with the random effects, and to assess effects of hypothesized/experimental random dimensions to cluster the observed data. The covariance structure model comprehends a large class of models with for instance the confirmatory factor model, the structural equation model, and the mixed effects model as special cases \cite{joreskog1970general}.
    
    \subsection*{Advantages}
	There are numerous advantages of modeling multi-way nested data with a structured covariance matrix in contrast to using random effect parameters. The structured covariance matrix can represent negative as well as positive within-cluster dependencies. This is a novel and important extension, since negative within-cluster differences can represent differences between clustered units. For instance, negative (positive) associations among clustered units can represent heterogeneity (homogeneity) in treatment effects between those receiving the same treatment. 
	
	Furthermore, the point of no association -- a zero covariance represents a random effect variance of zero -- is not a singularity under the BCSM. The covariance parameters can be negative, zero or positive, taking into account the positive-definiteness constraints on the covariance parameters. Thus, the BCSM does not require for instance mixture priors to include the singularities of the mixed effects model. This has the advantage that the covariance structure modeling approach does not have estimation and/or test issues when the covariance parameter approaches zero. This enables statistical testing of the significance of covariance parameters. For instance, testing the significance of the covariance between event times in the same treatment group represents testing overall non-inferiority between treatments.	Furthermore, each type of clustering is associated with a single covariance parameter, and the number of covariance parameters does not depend on the number of clusters. This makes the approach much more suitable for high-dimensional nested data in comparison to models that use a random effect for each type of clustering.  
	
	\subsection*{Contributions}
	Our BCSM for multi-way nested (categorical) data requires several technical innovations. First, expressions for the determinant and inverse of the structured covariance matrix are readily available from  \cite{searle1979dispersion}, \cite{lamotte1972notes}, and \cite{von2011inverse}. When the covariance parameters are positive, the structured covariance matrix is diagonally dominant and therefore positive definite. However, for negative covariance parameters this is not necessarily true. For balanced and some unbalanced designs, the necessary and sufficient conditions for a balanced design, and sufficient conditions for an unbalanced design for the covariance parameters are derived for which the structured covariance matrix is positive definite.    
	
	Second, conjugate shifted-inverse gamma prior distributions \cite{fox2017bayes} are proposed for the covariance parameters, which includes the positive-definite restrictions. It is shown that with novel  Helmert matrix transformations the structured covariance matrix can be diagonalized. This procedure is used to derive the analytical expressions for the posterior distributions of the covariance parameters. This result facilitates a \textit{Gibbs sampling algorithm}, and efficient sampling of covariance parameter values is demonstrated.     
	
	Third, for multi-way nested categorical data, a novel data augmentation (DA) algorithm is given. Expressions are given for the distribution of the latent (missing) data for any number of (nested) clusters and cluster sizes. The DA algorithm is computationally efficient, even when increasing the cluster size, since the inverse is explicitly known and each conditional covariance matrix is invariant across cluster members and clusters. This does not hold for an unrestricted covariance matrix, where the partitioning of the covariance matrix for each augmented latent variable leads to unique distributional components and requires computing the inverse of conditional covariance matrices. This leads to a significant increase in the computational burden of Markov chain Monte Carlo (MCMC) algorithms, when increasing the cluster size. Given the computational burden, MCMC-based algorithms for the multinomial probit model \cite{mcculloch1994exact,imai2005bayesian,jiao2015corrected,10.1214/20-BA1233} and the multivariate probit model \cite{chib1998analysis} are limited to (relatively) small cluster sizes. 
	
	Fourth, the BCSM is represented as a multivariate extension of the linear transformation model. This BCSM linear transformation model generalizes the normal frailty probit model for single-nested event times \cite{Wu2019normal}, and extends normal frailty (probit) survival models, and the multivariate linear transformation model of \cite{hothorn2019marginally}, by also allowing negative event time associations. Furthermore, the performance of the BCSM linear transformation model and the Gibbs sampler is shown on multi-nested event time data under type-II interval censoring, where the dimension of the covariance matrix is high (a block diagonal matrix of around 10,000 with main diagonal block sizes of around 3,300 observations). The study design represents a real-world scenario where the number of patients in each treatment group determines the dimension of the covariance matrix, which grows at the same rate as the cluster size.
	
	This paper is organized as follows: Section \ref{nesteddesigns} presents the BCSM for multi-way nested designs. Furthermore, the conditions are derived under which a multi-way nested covariance matrix is positive definite. In Section \ref{postcomp}, the posterior distributions are derived for the latent variables and covariance parameters under the BCSM. Section \ref{datadescr} introduces the BCSM linear transformation model for multi-way nested survival data and presents a Gibbs sampling algorithm. The performance of the Gibbs sampler is shown in a simulation study for two different nested designs in Section \ref{simulationstudies}. In Section \ref{application} our real-data application is presented in which the BCSM is used to analyze interval-censored clustered event time data from a three-armed multi-centre randomized clinical trial. Finally, Section \ref{discussion} is the conclusion with a discussion of the BCSM for nested designs.  
	
	\section{The BCSM for Nested Designs\label{nesteddesigns}}
	
	Following \cite{muthen1979structural} and \cite{bock1996high}, the dependence structure of multiple quantal variables can be described by $Q$ underlying factor variables (i.e., latent variables, random effects). In the nested design, the factor variables are independently normally distributed and nested within each other, where factor variable $q$ is nested within $q+1$ for $q=1,\ldots,Q-1$. A convenient representation of the BCSM is through underlying latent variables, denoted as $\mathbf{Z}$, also referred to as latent data, which are manifested through a threshold specification \cite{chib1998analysis}. The latent data specification is only necessary when the $\mathbf{Z}$ cannot be directly observed. Otherwise the $\mathbf{Z}$ is considered to be the observed outcome data. For the BCSM, given the factor variables, the implied dependence structure is limited to positive intra-cluster correlations, since the random effect variances are restricted to be positive.
	
	When integrating out the random effects, the BCSM has a covariance matrix representing dependencies implied by the random effects. Then, the latent variables in level $i$ of highest-level factor $Q$ are multivariate normally distributed -- without specifying a many-to-one mapping of the latent variables to the discrete observations -- according to
	\begin{eqnarray}
	\mathbf{Z}_i & \sim & N\left(\mathbf{X}_i\bm{\beta},\bm{\Sigma} \right),  \nonumber \\
	\bm{\Sigma} & = & \tau_0\mathbf{I} + \sum_{q=1}^Q\tau_q\mathbf{N}_q\mathbf{N}^\top_q, \label{generalsigma}
	\end{eqnarray}
	where $\tau_q$ ($q\ge1$) represents the intra-cluster correlation of factor $q$, and $\mathbf{N}_{q}$ is an incidence matrix representing the clustering of observations according to factor variable $q$. In the above, $\bI$ denotes the identity matrix, where we will often denote its dimension with a subscript in the remainder. For categorical observed data, without loss of generality, the covariance of factor 0 of the latent variables $\mathbf{Z}_i$ can be restricted $\tau_0=1$. The covariance matrix represents the nested structure and within-cluster dependencies are allowed to be negative through the covariance parameters. The covariance matrix $\bm{\Sigma}_i$ represents the dependence structure of the latent variables in level $i$ of highest-level factor $Q$, and we omit the index $i$ when only considering the latent variables of a single level of factor $Q$, as seen in Equation \eqref{generalsigma}.
	
	\subsection{Nested Design Notation}
		In an analysis of variance, the nested covariance matrix of an \textbf{unbalanced Q-way (random) nested design} has a certain structure. The matrix $\bm{\Sigma}$ represents the covariance matrix of the observations in level $i\in\{1,\dots,n_Q\}$ of factor $Q$. For $q<Q$, let $m_q$ denote the number of unique levels of factor $q$ in a level of factor $Q$, and $m_Q=1$. Let $n_{qj}$ denote the number of levels of factor $q$ nested in level $j$ of factor $q+1$. Due to the nested design, it follows that $m_q = \sum_j n_{qj}$. 
	Let $s_{qj}$ represent the number of observations (factor $0$) in level $j$ of factor $q$, and $s_{1j}=n_{0j}$. Then, the covariance matrix $\bm{\Sigma}$ is given by (see Remark \ref{Rem: lamotte})
	\begin{eqnarray}
	\bm{\Sigma}   = \tau_0 \mathbf{I}_{m_0} + \sum_{q=1}^{Q} \tau_q \left(\mathbf{I}_{m_q} \otimes \{\mathbf{J}_{s_{qj}}\}_j \right), \label{unbalancedsigma} 
    \end{eqnarray}
	where each $\bJ_{s_{qj}}$ is a square matrix of length $s_{qj}$ with all elements equal to one. Furthermore, the operator $\bI\otimes \{\bA_j\}_j$ denotes the matrix direct sum for a sequence of matrices $\{\bA_j\}_j$. The covariance matrix in Equation (\ref{unbalancedsigma}) is referred to as a \textit{nested unbalanced covariance matrix}. The covariance parameters $\btau$ can be identified when each factor contains a level with at least two lower levels nested in it. For $\bm 1_{s_{qj}}$, the all-ones vector in $\mathbb{R}^{s_{qj}}$, we have that \eqref{unbalancedsigma} equals \eqref{generalsigma} with $\bN_q = \bI_{m_q}\otimes \{\bm 1_{s_{qj}}\}_j$. 

	For the \textbf{balanced $Q$-way nested design}, we have $n_{qj}=n_q$ for all $q$ i.e., the number of unique levels, $n_{q}$, of each factor variable $q$ ($q=0,\ldots,Q-1$), is equal across levels of higher factor variable $q+1$. The nesting information is represented by $\mathbf{n} \in \mathbb{N}^{Q}$, which is referred to as the {\it nesting} vector. The number of observations in each unique level of factor $q$ is denoted $s_q$, which is the cumulative product of $\mathbf{n}$ up to $q-1$, $s_q = \prod_{r=0}^{q-1} n_{r}$, where $s_0=1$. For a level of the highest factor $Q$, in the balanced design the number of  nested levels of factor $q$ can be expressed as $m_q = s_Q/s_q$, where $s_Q$ represents the total number of observations in each level of factor $Q$. Then, for the balanced nested design, the covariance matrix is represented by
	\begin{eqnarray}
	\bm{\Sigma} &=& \tau_0\mathbf{I}_{s_{Q}} + \sum_{q=1}^Q \tau_q(\mathbf{I}_{m_q}\otimes \mathbf{J}_{s_q}),\label{balancedsigma} 
	\end{eqnarray}
	 where $\otimes$ denotes the Kronecker product. A matrix of the form of Equation (\ref{balancedsigma}) is referred to as a {\it nested balanced covariance matrix}. 
	 \begin{Remark}[Construction of the multi-way nested covariance matrix]\label{Rem: lamotte}
	 The covariance matrix $\bSigma$ in Equation \eqref{unbalancedsigma} can be recursively constructed according to \cite{lamotte1972notes}. To this end, let the design matrix $\tilde{\mathbf{N}}_{q}$ denote membership of factor $q$ in a level of factor $q+1$. Using the notation of Equation \eqref{unbalancedsigma}, the design matrix $\tilde{\mathbf{N}}_{q}$ is given by 
	\begin{eqnarray}
	\tilde{\mathbf{N}}_{q} & = &  \mathbf{I}_{m_{q}} \otimes \{\mathbf{1}_{n_{qj}}\}_j.  \nonumber 
	\end{eqnarray}
	 Let $\tilde{{\Sigma}}_Q = \tau_Q$, and let $\tilde{\bm{\Sigma}}_{q}$ represent the nested dependence structure induced by the factors $Q$ up to and including $q$. Then, $\bSigma = \tilde{\bSigma}_0$, where for all $q<Q$
	\begin{eqnarray}
	\tilde{\bm{\Sigma}}_{q} & = & \tau_q \mathbf{I}_{m_q} + \tilde{\mathbf{N}}_{q} \tilde{\bm{\Sigma}}_{q+1} \tilde{\mathbf{N}}^\top_{q}. \nonumber 
	\end{eqnarray}
	Furthermore, $\bSigma$ equals the covariance matrix in \eqref{generalsigma} for $\bN_{q}= \bI_{m_q}\otimes \{\bm 1_{s_{qj}}\}_j$.
	\end{Remark}

	\subsection{Parameter Restrictions for the Nested Covariance Matrix}\label{posdefprop}
	The inverse of the nested covariance matrix in Equations \eqref{unbalancedsigma} and \eqref{balancedsigma} 
	plays a crucial role when estimating parameters and computing the covariance matrix of estimators
for mixed effect models. In our Bayesian approach, the inverse and the positive-definiteness constraints are important to specify the posterior distributions for the covariance parameters and the distributions of the latent variables. Therefore, the inverse has been studied extensively.
	
	Formulas for the eigenvalues, determinant and the inverse under a balanced design are given by \cite{searle1979dispersion}. \cite{lamotte1972notes} defined recursive procedures for the determinant and the inverse for any nested unbalanced classification. Explicit expressions for the inverse for unbalanced nested designs are given by \cite{stepniak1995inverting}. \cite{wansbeek1982simple} presents the spectral decomposition of a balanced covariance matrix to obtain the eigenvalues and inverse in a straightforward way. The coefficients of the expression for the (balanced) inverse of \cite{searle1979dispersion} are explicitly given by \cite{jiang2004dispersion} and to a broader extent by \cite{von2011inverse}.   
	
	Positive-definite constraints for the covariance parameters are only relevant when the covariance parameters are not restricted to be positive. Traditionally, each covariance parameter represents a (positive) random effect variance, which is restricted to be positive. This makes the covariance matrix diagonally dominant and hence automatically positive definite. In our setup for the BCSM, the covariance parameters are not restricted to be positive, hence the positive definite property of the covariance matrix does not necessarily hold and conditions for positive definiteness must be derived.
	
	The following theorem gives a necessary and sufficient condition under which a nested balanced covariance matrix is positive definite, with covariance parameters allowed to be negative.
	\begin{theorem}\label{constraintbalanced}
		The nested balanced covariance matrix in Equation \eqref{balancedsigma} is positive definite if and only if $\tau_0>0$ and 
		\begin{eqnarray}
		\tau_q & > & \frac{-\left(\tau_0 + \sum_{r=1}^{q-1}s_{r}\tau_{r}\right)}{s_q}\;\;\;\;\forall q\in\{1,\dots, Q\}.
		\end{eqnarray}
	\end{theorem}
	\begin{proof}
		The unique eigenvalues of a nested balanced covariance design matrix are given by (see e.g., Eq. 3.10 in \cite{searle1979dispersion})
		\begin{eqnarray}
		\tau_0+\sum_{r=1}^qs_r\tau_r\;\;\;\;\forall q\in\{0,\dots, Q\}.
		\end{eqnarray}
		The eigenvalues of $\bSigma$ can also be determined recursively by observing that 
		\begin{eqnarray}
		\bSigma = \underbrace{\tau_0\bI_{s_Q} + \sum_{q=1}^{Q-1}\tau_q(\bI_{m_q}\otimes \bJ_{s_q})}_{\bSigma_{Q-1}} + \tau_{Q}\bm 1_{s_Q} \bm 1_{s_Q}^\top.
		\end{eqnarray}
		The vector $\bm 1_{s_Q}^\top$ is a left-eigenvector of $\bSigma_{Q-1}$ with eigenvalue $\tau_0+\sum_{q=1}^{Q-1}s_q\tau_q$. Following \cite{ding2007eigenvalue}, the eigenvalues of $\bSigma$ are the eigenvalues of $\bSigma_{Q-1}$ but with one of the eigenvalues equal to $\tau_0+\sum_{q=1}^{Q-1}\tau_q s_q$ replaced by $\tau_0+\sum_{q=1}^{Q}\tau_qs_q$. The result now follows by induction and the fact that a matrix is positive definite if and only if all the eigenvalues are positive.
	\end{proof}
	
	 No general results are known for the spectral decomposition of an unbalanced nested covariance matrix \cite{wansbeek1982another}. However, the nested unbalanced covariance matrix defined in Equation \eqref{unbalancedsigma} can be regarded as a principal submatrix of a (maximum) nested balanced covariance matrix. Let $\bar{n}_q = \max_{j} n_{qj}$ denote the maximum number of unique levels of factor $q$ across levels $j$ of factor $q+1$, and also across levels $i$ of factor $Q$ to balance the covariance matrix for all components $\mathbf{Z}_i$. Then, $\bar{s}_q = \prod_{r=0}^{q-1} \bar{n}_r$ represents the (balanced) number of observations in each level of factor $q$ and $\bar{m}_q = \bar{s}_{Q}/\bar{s}_q$ the number of nested levels in layer $q$ for this maximum balanced matrix. It follows that each $\bN_{q}$ is a (row) submatrix of $\bI_{\bar{m}_q}\otimes \bm 1_{\bar{n}_q}$ and hence $\bm{\Sigma}$ from Equation \eqref{unbalancedsigma} is a principal submatrix of the maximum nested balanced covariance matrix: 
	\begin{equation}
	\bSigma_b = \tau_0\bI_{\bar{s}_{Q}} + \sum_{q=1}^Q \tau_q(\bI_{\bar{m}_q}\otimes \bJ_{\bar{s}_q}). \label{balancedversion}
	\end{equation}
	It holds that each $\bm{\Sigma}$ is positive definite if $\bm{\Sigma}_b$ is positive definite, since each $\bm{\Sigma}$ is a principal submatrix of $\bm{\Sigma}_b$.  
	
	Alternatively, consider the minimum balanced principal submatrix of the nested unbalanced covariance matrix in Equation \eqref{unbalancedsigma}, with for each factor $q$ the number of nested levels in the balanced case equal to the minimum number of nested levels in the unbalanced case. Let $\ubar{s}_q$ represent the (balanced) number of observations in each level of factor $q$ based on the minimum number of unique levels of factor $q$ across levels $j$ and $i$ of factor $q+1$ and $Q$, respectively. When $\bm{\Sigma}$ is positive definite, any principal submatrix must be positive definite which includes the minimum balanced principal submatrix. 
	
	From the above discussion, it is seen that the following necessary condition and sufficient condition can be given for the covariance parameters to ensure that the nested unbalanced covariance matrix is positive definite.  
	\begin{corollary}\label{corolunbalanced}
		The nested unbalanced covariance matrix $\bm{\Sigma}$ is positive definite if $\tau_0>0$ and 
		\begin{eqnarray}\label{unbalancedrestriction2} 
		\tau_q & > & \frac{-\left(\tau_0 + \sum_{r=1}^{q-1}\bar{s}_{r}\tau_{r}\right)}{\bar{s}_q} \;\;\;\;\forall q. 
		\end{eqnarray}
		Furthermore, $\bm{\Sigma}$ is positive definite only if $\tau_0>0$ and 
		\begin{eqnarray}\label{unbalancedrestriction2b} 
		\tau_q & > & \frac{-\left(\tau_0 + \sum_{r=1}^{q-1}\ubar{s}_{r}\tau_{r}\right)}{\ubar{s}_q} \;\;\;\;\forall q. 
		\end{eqnarray}
	\end{corollary}
	
	A special case occurs when the parameter space for $\bm{\tau}$ remains the same under the (extended) balanced version $\bm{\Sigma}_b$. This occurs, for instance, when only the number of levels of factor $Q-1$ varies across levels of highest-level factor $Q$, a {\it top-layer unbalanced nested design}. Then, the constraint in Equation \eqref{unbalancedrestriction2} also becomes a {\it necessary} condition. Let $s_{Qi}$ represent the number of observations in level $i$ of the highest-level factor $Q$ and let $s_q$ represent the number of observations in each level of layer $q<Q$ for the balanced part of the covariance structure.  Then, according to Theorem \ref{constraintbalanced}, it holds that \textit{each} $\bSigma_i$ is positive definite if and only if $\tau_0>0$ and
	\begin{eqnarray}
	\tau_q & > & \frac{-\left(\tau_0 + \sum_{r=1}^{q-1}s_{r}\tau_{r}\right)}{s_q} \;\;\;\;\forall q<Q, \label{restrictunbalanced1} \\
	\tau_Q & > & \frac{-\left(\tau_0 + \sum_{r=1}^{Q-1}s_{r}\tau_{r}\right)}{s_{Qi}}\;\;\;\;\forall i \implies \tau_Q> \frac{-\left(\tau_0 + \sum_{r=1}^{Q-1}s_{r}\tau_{r}\right)}{\bar{s}_Q} . \label{restrictunbalanced2}
	\end{eqnarray} 
	Hence, the parameter space of $\bm \tau$ does not change when replacing each $\bSigma_i$ by $\bSigma_b$. This occurs in our real data example, where only the number of patients (first-level factor), each with three event types, varies across treatment groups (second-level factor). A sampling procedure is developed in which each main-diagonal block $\bSigma_i$ of the nested unbalanced covariance matrix is artificially augmented to a nested balanced covariance matrix $\bSigma_b$. This procedure transforms the posterior analysis of the covariance vector $\bm \tau$ to a balanced situation, which greatly simplifies the analysis.
	
	\subsubsection*{Relation to ML Estimation} Corollary \ref{corolunbalanced} represents a {\it sufficient} condition (Equation \eqref{unbalancedrestriction2}) for a positive definite nested unbalanced covariance matrix. Hence, in the unbalanced design it is possible that certain covariance vectors $\bm \tau$ are allowed for which $\bm{\Sigma}_b$ is non-positive definite. When constraints on the negative part of the parameter space are less restrictive under the unbalanced design than under the (extended) balanced design, constraints are too restrictive -- when estimating under the (extended) balanced design -- resulting in biased estimates. Note that the parameter space defined in Corollary \ref{corolunbalanced} includes the non-negative constraints on the (co)variance components and therefore still extends the usually considered parameter space under maximum likelihood (ML) estimation. In general, for ML estimation in variance component models, the likelihood function is maximized over the positive space of the variance components to obtain the ML estimators of variance components \cite{CorbeilSearle1976}. \cite{Harville1977} discussed procedures that constrain algorithmic iterates to non-negative values. For instance, Henderson's iterative algorithm for computing ML and REML estimates of variance components ensure non-negative values for the variance components at any point when the algorithm starts with strictly positive values. Under the constraint in Equation \eqref{unbalancedrestriction2} of Corollary \ref{corolunbalanced}, it is still possible to explore negative correlation among clustered observations, which is not possible under most regular ML estimation methods. Furthermore, the point zero is not a boundary value of the parameter space, which facilitates statistical testing whether a covariance is positive, zero, or negative. Classical test approaches for testing the hypothesis of homogeneity (zero random effect variance) against heterogeneity (positive random effect variance) break down, since under the null hypothesis the variance parameter is at the boundary of the parameter space	\cite{pauler1999bayes,claeskens2008one}.
	
	\section{Posterior Computation~\label{postcomp} }
	An iterative procedure is derived to determine the full conditional distribution of the marginals of $\bZ_i$ in closed form under a nested balanced design. Furthermore, a class of conjugate priors for the covariance parameters $\btau$ is determined. Both results facilitate a Gibbs sampling algorithm based on censored data, where observations consist of sets $\bOmega_i$ such that $\bZ_i\in\bOmega_i$ for all $i$. For an unbalanced nested design, a data augmentation procedure is proposed to create an artificially balanced nested design (see Remark \ref{rem:unbalancedZ}). Then, posterior distributions for the balanced design can also be applied to unbalanced nested data. 
	
	\subsection{Conditional Distribution of Latent Variables} 
	A recursive procedure is given to derive the full conditional distribution of the marginals of $\mathbf{Z}_i$ under a balanced nested design. This requires an analytical expression for the inverse of the nested balanced covariance matrix:
	\begin{lemma}\label{lemmainvbalanced}
		Let $v_q = \tau_0 + \sum_{r=1}^qs_r\tau_r$ and 
		\begin{equation}\rho_q =  \frac{-\tau_q}{v_{q}v_{q-1}}\;\;\;\; \forall q\geq 1.  \label{rhodef}\end{equation}
		The inverse of $\bSigma$ in Equation \eqref{balancedsigma} is given by
		\begin{eqnarray}
		\bSigma^{-1} &=& \frac{1}{\tau_0}\bI_{s_{Q}} + \sum_{q=1}^Q \rho_q (\bI_{m_q}\otimes \bJ_{s_q}).\label{invmatbalanced}
		\end{eqnarray}
	\end{lemma}
	\begin{proof}
		This result can be verified by  observing that
		\begin{eqnarray}
		\frac{1}{\tau_0}+\sum_{r=1}^q s_{r}\rho_{r} &=& \frac{1}{v_{q}}\;\;\;\; \forall q\label{propertyrho}
		\end{eqnarray}
		which implies that $\brho$ satisfies Eq. 4.7 in \cite{searle1979dispersion}, noting that $v_q,\tau_0$ are the eigenvalues of $\bSigma$.
	\end{proof}
	
	For a matrix $\bA$ this makes it relatively straightforward to evaluate the expression
	\begin{equation}
\bA^\top\bSigma^{-1}\bA =  \frac{1}{\tau_0}\bA^\top\bA + \sum_{q=1}^Q\rho_q\;\bA^\top(\bI_{m_q}\otimes \bJ_{s_q})\bA.\label{opinverse}
	\end{equation}
	Evaluation of such an expression is required for the computation of the full conditional distribution of regression parameters. This result will be used to efficiently perform posterior inference for the spline and regression parameters under the BCSM introduced in Section \ref{datadescr}. Furthermore, the computational demand of the Gibbs sampling algorithm is reduced, since the (low-dimensional) matrix products $\bA^\top(\bI_{m_q}\otimes \bJ_{s_q})\bA$ on the right-hand side of \eqref{opinverse} stay fixed over iterations and can hence be stored before running the algorithm.
	
	Next, it is shown that the conditional distribution of the latent variables in a level of factor $Q-1$ given all other observations in a level of factor $Q$ is again a multivariate normal distribution with a nested balanced covariance matrix, but now with $Q-1$ factors. The proof is given in Appendix \ref{prooftheorem32}. This result can be recursively applied in order to derive the marginal full conditional distribution of the latent variables, which is particularly useful when sampling under a restricted support (e.g, when dealing with categorical or event time data).  
	
	\begin{theorem}\label{thmZsamp}
		Let $\mathbf{Z}_i\sim N(\bm{\mu}_i,\bm{\Sigma})$ with $\bm{\Sigma}$ positive definite as defined in \eqref{balancedsigma}. Let $\mathbf{Z}_{ij}$ denote the observations in the $j$-th cluster of factor $Q-1$. Under full knowledge of $\bmu_i$ and $\bSigma$, the $\mathbf{Z}_{ij}$ (with mean $\bm \mu_{ij}$) given remaining observations $\mathbf{Z}_{i(-j)}$ (with mean $\bm\mu_{i(-j)}$) is multivariate normally distributed with a balanced nested covariance matrix 
		\begin{eqnarray*}
			\mathbf{Z}_{ij} \mid \mathbf{Z}_{i(-j)} & \sim & N(\bm \theta_{ij},\; \bSigma_{Q-1}) 
		\end{eqnarray*}
		where
		\begin{eqnarray*}
			\bm \theta_{ij} &=& \bm\mu_{ij} +  c_{ij}\bm 1_{s_{Q-1}},\\
			\bSigma_{Q-1} &=& \tau_0\bI_{s_{Q-1}} + \sum_{q=1}^{Q-1}\tau_q (\bI_{m'_q}\otimes \bJ_{s_q}) + \tau_Q(1-f_Q)\bJ_{s_{Q-1}},
		\end{eqnarray*}
		and, letting $\bar{\nu}$ denote the average over entries of a vector ${\bm \nu}$,
		\begin{eqnarray*}
			c_{ij} &=& f_Q\left(\bar{Z}_{i(-j)}-\bar{ \mu}_{i(-j)}\right),  \, f_{Q} = \frac{u_{Q}\tau_Q}{v_{Q-1} +u_Q\tau_Q},  \\
			u_Q &=& (m_{Q-1}-1)s_{Q-1}, \, v_{Q-1} = \tau_0+\sum_{q=1}^{Q-1}s_q\tau_q,\; m'_q = m_q/n_{Q-1}.
		\end{eqnarray*}
	\end{theorem}\phantom{.}\\

	\begin{Remark}[Data Augmentation Procedure] \label{rem:unbalancedZ}
	For an unbalanced nested design, latent data vector $\bZ_i$ with an unbalanced nested covariance matrix \eqref{unbalancedsigma} is augmented to a larger vector  $\bZ_i^\text{tot}$ which has a balanced nested covariance matrix according to Equation \eqref{balancedversion} and a known mean vector agreeing with the one of $\bZ_i$. If the covariance matrix is positive definite given $\btau$, each $\bZ_i^\text{tot}$ can be sampled in a Gibbs sampling procedure according to Theorem~\ref{thmZsamp}, and the stationary marginal distribution of the subvector $\bZ_i$ equals the posterior distribution of $\bZ_i.$ When dealing with categorical or event time data, the procedure still works with the possible inclusion of a restriction on the support of the latent vector $\bZ_i$.
	Introducing additional latent variables can increase the autocorrelation of sampled values. Note that for a top-layer unbalanced nested design, Theorem \ref{thmZsamp} can be applied to each independent block, and hence no additional latent variables have to be sampled for this step in a Gibbs sampler. 
	\end{Remark}

	\subsection{Posterior Distribution of the Covariance Parameters}
	For the nested balanced covariance matrix \eqref{balancedsigma}, the posterior distribution of the (possibly negative) covariance parameters is derived. This leads to a novel Gibbs sampling procedure (see Appendix \ref{Gibbssampler}).
	The posterior distributions are based on a transformation of the latent variables, which are rescaled to have mean zero
	\begin{eqnarray}
	\mathbf{V}_i &=& \mathbf{Z}_i - \bmu_i, \nonumber \\
	\mathbf{V}_i & \sim & N\left(\bm 0,\; \bm{\Sigma}\right) \nonumber .
	\end{eqnarray}
	Given $\bmu_i$, the vectors $\mathbf{V}_i$ are independently distributed and contain the relevant data information about the covariance parameters $\bm{\tau}$.
	
	
	Under a balanced nested design, orthonormal Helmert transformation matrices (e.g., \cite{gentle2012numerical}), $\mathbf{H}_n$ in $\mathbb{R}^{n\times n}$, can be used to diagonalize the covariance matrix. The Helmert transformations that diagonalize $\bm{\Sigma}$ are operated on the rescaled latent variables $\mathbf{V}_i$, to obtain sufficient statistics for each covariance parameter and to construct posterior distributions. 
	
	A product of Helmert matrices is defined that diagonalize $\bm{\Sigma}$. The components of the covariance matrix in Equation \eqref{balancedsigma} can be represented as a Kronecker product of smaller all-ones ($J$-)matrices and identity matrices.\\ Letting $\bigotimes_{i=1}^n \bA_i=\bA_1\otimes \bA_2\otimes\cdots\otimes \bA_n$ for a sequence of matrices $(\bA_i)_{i=1}^n$, it holds that
	\begin{eqnarray}
	\bI_{m_q}\otimes \bJ_{s_q} &=& \left(\bigotimes_{r= Q-1}^{q}\bI_{n_{r}}\right) \otimes \left(\bigotimes_{r = q-1}^{0} \bJ_{n_{r}}\right).\label{expressioncrossprod}
	\end{eqnarray}
	Subsequently, it is shown that both $\bI_{n_q}$ and $\bJ_{n_q}$ can be diagonalized by $\bH_{n_q}$. 
	
	When viewed as an operator on vectors, the Helmert matrix decomposes a vector in a term which is proportional to its mean (first element) and terms that represent the deviations. For a constant vector, these deviations are zero, and $\mathbf{H}_{n_q} \mathbf{1}_{n_q} = \sqrt{n}_q\mathbf{u}_{1n_q}$, with $\bu_{1n_q}$ the first unit vector in $\mathbb{R}^{n_q}.$ Hence, when multiplying the $J$-matrix on the left and right with the Helmert matrix, it can be seen that the result is a diagonal matrix
	\begin{eqnarray}
	\mathbf{H}_{n_q}\bJ_{n_q}\bH_{n_q}^\top &=& (\bH_{n_q}\bm 1_{n_q})(\bm 1_{n_q}^\top\bH_{n_q}^\top)=n_q\bu_{1n_q}\bu_{1n_q}^\top= n_q\bK_{n_q}
	\end{eqnarray}
	where $\bK_{n_q}\in\mathbb{R}^{n_q\times n_q}$ represents the single-entry matrix with a one at position $(1,1)$ and all other elements zero. When operating $\mathbf{H}_{n_q}$ on the identity matrix, the result is also a diagonal matrix, since the Helmert matrix is orthonormal such that $\bH_{n_q}\bI_{n_q}\mathbf{H}_{n_q}^\top = \bI_{n_q}$.
	
	
	It follows that for each $q$ the Helmert matrix $\bH_{n_q}$ diagonalizes $\bI_{n_q}$ and $\bJ_{n_q}$. Therefore, by Equation \eqref{expressioncrossprod},  $\bm{\Sigma}$ can be diagonalized with a Kronecker product of Helmert matrices
	\begin{eqnarray}
	\mathbf{H} = \bigotimes_{q=Q-1}^0 \bH_{n_q},
	\end{eqnarray}
	taking into account that the $n_q$ levels of factor $q$ are nested in each level of factor $q+1$. 
	
	Theorem \ref{posteriorstau} below shows that for the balanced nested design, the shifted-inverse gamma distribution is a conjugate prior for each covariance parameter, when conditioning the posterior distribution on $\bZ,\bmu$ and the other covariance parameters.	The shift parameter depends on covariance parameters of the nested factors. The scale parameter is constructed from an idempotent projection matrix operated on the multivariate normally distributed (latent) variables. As in \cite{fox2017bayes}, the conjugate shifted-inverse gamma prior for the covariance parameters  has support $(-\sigma,\infty)$ where it is defined as 
	\begin{equation}
	\text{shifted-}\mathcal{IG}(x;\;a,\;b,\;\sigma) = \frac{b^a}{\Gamma(a)}(x + \sigma)^{-(a+1)}\exp(-b/(x+\sigma)).
	\end{equation}
This leads to the following result, for which the proof is given in Appendix \ref{prooftheorem33}.
	
	\begin{theorem}\label{posteriorstau}
		Let $\bV_i\sim N(\bm 0,\;\bSigma)$, where $\bSigma$ is a nested balanced covariance matrix (Equation  (\ref{balancedsigma})). Assume the following shifted inverse-gamma priors for the covariance parameters,
		\begin{eqnarray*}
		\tau_0 &\sim &\mathcal{IG}(\alpha_{\tau_0},\beta_{\tau_0}), \\
		\tau_q\mid \btau_{<q} &\sim& \text{shifted-}\mathcal{IG}\left(\alpha_{\tau_q},\beta_{\tau_q},\left(\tau_0 + \sum_{r=1}^{q-1}s_r\tau_r\right)/s_q\right),\;\;\;\; \forall q\in\{1,\dots, Q\},
		\end{eqnarray*}
		where $\btau_{<q} = [\tau_0,\dots,\tau_{q-1}]$.
		Then, each covariance parameter $\tau_q$ has a shifted inverse-gamma posterior distribution
		\begin{eqnarray*}
		\tau_0\mid\bV &\sim &\mathcal{IG}\left(\alpha_{\tau_0} + n_Qp_0/2,\beta_{\tau_0} + \frac{\sum_{i=1}^{n_Q}S_{i0}^2}{2}\right),\\
		\tau_q\mid(\bV,\btau_{<q})&\sim& \text{shifted-}\mathcal{IG}\left(\alpha_{\tau_q} + n_Qp_q/2,\;\beta_{\tau_q}+\frac{\sum_{i=1}^{n_Q}S_{iq}^2}{2}, \;\left(\tau_0+\sum_{r=1}^{q-1}s_r\tau_r\right)/s_q\right),
		\end{eqnarray*}
		for $q\in\{1,\dots, Q\}$. The scale parameter components are defined by the sum-of-squares  
		\begin{eqnarray}
		S_{iq}^2 = \|\bM_q\bH\bV_i\|^2/s_q, \nonumber 
		\end{eqnarray}
		using the idempotent subspace projection matrices
		\begin{align*}
		\bM_q &=  \left(\bm \bI_{m_{q}}\otimes  \bK_{s_{q}}\right)-
		\left(\bI_{m_{q+1}}\otimes \bK_{s_{q+1}}\right),\;\;\;\; \forall q\in\{0,\dots, Q-1\},\\
		\bM_Q&= \bI_{m_Q}\otimes \bK_{s_Q}.
		\end{align*}
		The shape parameters $p_q$ are represented by
		\begin{eqnarray}
		p_q &=&  \text{tr}(\bM_q) = \begin{cases}m_q - m_{q+1} ,\;\;\;\;&\text{if $q<Q$},  \\
		1,\;\;\;\;&\text{else.}\end{cases} \nonumber 
		\end{eqnarray}
		
		
	\end{theorem}\phantom{.}\\

	Using equivalence of the inner product and the trace of the outer product, as well as invariance under cyclic permutations of the trace, we can write for $q<Q$
	
	$$S_{iq}^2 = \sum_{\bell_{q:}\in\mathcal{L}_{q:}}(\bar{V}_{i\bell_{q:}} - \bar{V}_{i\bell_{(q+1):}})^2,\;\;\;\; S_{iQ}^2 = \bar{V}_i^2,$$
	where $\mathcal{L}_{q:}$
contains vectors denoting nested levels from level $q$ up to level $Q-1$ and $\bar{V}_{i\bell_{q:}}$ is the mean of the observations in the level of factor $q$ denoted by $\bell_{q:}=[\ell_q,\ell_{q+1},\dots, \ell_{Q-1}]$, where each $\ell_q\leq n_q$. When multiplied with a factor $s_q$, these sum of squares agree with the usual sum of squares seen in an analysis of variance of a random nested design \cite{searle1992variance}.

	\begin{Remark}[{Unbalanced Nested Design\label{covunbalanced}}]
	For an unbalanced design, larger clusters contain more information about the covariance parameters than smaller ones. Due to this imbalance in information, no convenient conjugate prior distribution is available for the covariance parameters for unbalanced designs. However, for some unbalanced nested designs the parameter space for $\bm \tau$ remains the same under an artificially (maximum) balanced version (see Equation \eqref{balancedversion}). In that case, the nested unbalanced covariance matrix is considered to be a principal submatrix of a maximum balanced covariance matrix. A data augmentation step can be defined to artificially balance the data matrix such that the covariance parameters can be sampled according to the posterior distributions defined in Theorem \ref{posteriorstau}. The augmentation procedure is valid as long as the marginal distribution of the covariance parameters remains the same, and the joint distribution of the parameters and observed data can be obtained from the joint distribution of the extended set\label{valid_DA}. We will see in Section \ref{datadescr} that for a top-layer unbalanced nested design, it suffices to complete the sample means to those under a balanced design. As this corresponds to (parallelized) univariate normal sampling, this can be done very efficiently.
	
	Constraints on the parameter space under an unbalanced design -- this concerns the negative part of the parameter space -- can be less restrictive than under a balanced design. Then, the parameter space described by the posterior distribution for the covariance parameters derived under an artificially (maximum) balanced design is incorrectly constrained. Note that the artificial balancing procedure will always work when assuming a positive covariance parameter.   
	\end{Remark}
	
	\section{Application to Interval-Censored Nested Survival Data\label{datadescr}}
	This section builds upon the results of Section \ref{postcomp} to construct a Gibbs sampler for the BCSM linear transformation model for multi-way nested event time data under type-II interval censoring. This BCSM (survival) model is a multivariate extension of the linear transformation model in which the normal cumulative distribution function is used to describe clustered event times. The model extends normal frailty probit survival models by also allowing negative event time associations. 
	
	The design of a real data study is followed \cite{von2016very}, where patients were assigned to $n_{2}$ treatment groups (factor $q=2$). Furthermore, $n_{1i}$ patients (factor $q=1$) were assigned to treatment group $i$, and event times of $n_0$ different event types were observed per patient. The dependence structure is represented by a two-way nested unbalanced design. For each event time, only the day at which the event time occurred is known, hence the event times are type-II interval censored.
	
	The event times $\bt_i$ of treatment group $i$ are assumed to be described by the BCSM linear transformation model
	\begin{align}
	\bh_i(\bt_i) &= -\bX_i\bm\beta + \bE_i,\label{LTMmod}\\
	\bE_i&\sim N(\bm 0, \bSigma_i),\nonumber\\
	\bSigma_i & =  \bI_{n_{0}n_{1i}} + \tau_1\left(\bI_{n_{1i}} \otimes \bJ_{n_{0}}\right) + \tau_2 \bJ_{n_{0}n_{1i}},\nonumber
	\end{align}
	where the structured covariance matrix represents the nesting of event times in patients, who are nested in treatment group $i$. Notice that only the  first factor (number of patients, factor $Q-1$) is unbalanced in the nested unbalanced covariance matrix. Furthermore, $\tau_0=1$ is enforced for identification purposes, notice that this does not change the result of Theorem \ref{posteriorstau} as the posterior distributions of the other covariance parameters are given conditional on $\tau_0$.
	
	The transformation $\bh_i$ is such that $h_{ijk}$ is an increasing continuous function with $h_{ijk}(0)=-\infty$ and $h_{ijk}(\infty)=\infty$ for event time $k$ of patient $j$ in treatment group $i$. Next, for each event time $t_{ijk}$ we only observe an interval $\tilde{\Omega}_{ijk}=[L_{ijk}, R_{ijk})$ such that $t_{ijk}\in\tilde{\Omega}_{ijk}$, and it is allowed that $L_{ijk}=0$ or $R_{ijk}=\infty$. We assume non-informative censoring, meaning $$P(\bt_i\leq \bs|\bL_i=\bell_i, \bR_i=\br_i, \bX_i ) = P(\bt_i\leq \bs|\bell_i\leq \bt_i<\br_i, \bX_i).$$ The baseline function $\bh_i$ is modeled using monotone regression splines \cite{lin2010semiparametric}, where the $h_{ijk}$ are equal to the same translated linear combination of integrated splines parametrized by a vector $\bgamma$. The transformation applied to $\bt_i$ resulting from a given vector $\bgamma$ is denoted by $\bh_i(\cdot\mid\bgamma)$.
	
	Latent variables, $\bZ_i$, are introduced for posterior computation, which are linked to the interval restrictions on the event times, $\bL_i\leq \bt_i\leq\bR_i$, through a many-to-one mapping. Following the augmentation design of \cite{lin2010semiparametric} for the univariate probit model, let $s_{ijk}$ equal $L_{ijk}$ if $L_{ijk}>0$, and equal $R_{ijk}$ otherwise, and let $\bZ_i = \bh_i(\bs_i) - \bh_i(\bt_i)$. Then, the latent variables are multivariate normally distributed  
	\begin{eqnarray}
	\mathbf{Z}_i &  \sim & N\left(\bh_i(\mathbf{s}_{i})+\mathbf{X}_{i}\bm{\beta},\;\mathbf{\Sigma}_{i}\right), \label{survmodelintvcens} 
	\end{eqnarray}
	with the interval restriction $Z_{ijk}\in\Omega_{ijk}$
	\begin{equation}\Omega_{ijk} = \begin{cases}
	(h_{ijk}(L_{ijk}) - h_{ijk}(R_{ijk}),0]\;\;\;\;&\text{if }L_{ijk}>0 \text{ and }R_{ijk}<\infty, \\
	(0,\infty)&\text{if } L_{ijk} = 0,\\
	(-\infty,0]&\text{if } R_{ijk} = \infty.
	\end{cases}
	\label{omegadef}
	\end{equation}
    The sign of the covariance between clustered event times is equal to the sign of the covariance between the corresponding latent variables, since the latent variables are defined as a nondecreasing, non-constant function $h(.)$ of the event times (for the proof see Appendix \ref{covarianceproof}).~\label{statement_preservation}
	
	Let $\mathbb{I}(\cdot)$ denote the indicator function. From \eqref{survmodelintvcens}, it is seen that the augmented data likelihood becomes
	\begin{align}
	&\hspace{-5mm}\mathcal{L}\left(\bgamma,\bm{\beta},\btau \mid \mathbf{Z},\mathbf{X},\mathbf{L},\mathbf{R}\right)=\prod_{i=1}^{n_2}\phi\left(\mathbf{Z}_i\mid\bh_i\left({\mathbf{s}}_i\mid\bgamma\right)+\mathbf{X}_i\bm{\beta},\; \mathbf{\Sigma}_i\right) \mathbb{I}\left(\mathbf{Z}_{i} \in \bm{\Omega}_{i}\right).\label{auglik}
	\end{align}
	The (observed-data) likelihood cannot be factorized as a product of marginal components representing the likelihood contributions of the left, right and interval censored event times, since the event times are not independently distributed. Following the restrictions in Equations \eqref{restrictunbalanced1} and \eqref{restrictunbalanced2} for a nested design, the covariance matrix $\bSigma_i$ is positive definite for all $i$, under the following restrictions for the covariance parameters:
	\begin{eqnarray}
	\tau_1 & > & -1/n_0, \label{rhoconstraint}\\
	\tau_2 & > & \frac{-1-\tau_1n_0}{n_0\max_i n_{1i}}. \label{tauconstraint}
	\end{eqnarray}
	For these restrictions, clustered event times can be negatively, positively or zero correlated, for each type of cluster. 
	
	\subsection{Data Augmentation Gibbs Sampler \label{sect:GS}}
	Posterior computation is performed using a Gibbs sampling algorithm. The algorithm consists of four sampling steps: sampling of latent variables, covariate effects, spline coefficients, and covariance parameters. The full conditional distributions of covariate parameters $\bbeta$ and spline parameters $\bgamma$ are relatively straightforward, and they are given in Appendix \ref{Gibbssampler}.   
	
	The sampling of the latent variables and covariance parameters require a specific approach, to avoid computing the inverse of the nested covariance matrix, and to sample directly from the posterior distributions. The posterior distributions are derived using the results from Theorem \ref{thmZsamp} and Theorem \ref{posteriorstau}.

\subsubsection*{Sampling Latent Variables.~\label{sampleZsection}} 

The full conditional distribution is derived for the marginals of $\bZ_i$ defined in Equation \eqref{survmodelintvcens}. First, the full conditional distribution is derived of the latent variables of patient $j$ in treatment group $i$, $\bZ_{ij}$, given the latent variables of the other patients in treatment group  $i$, $\bZ_{i(-j)}$. Let $\bm \mu_{i} = \tilde{h}({\bs}_i)+\bX_i\bm \beta$, it follows from Theorem \ref{thmZsamp} and the likelihood \eqref{auglik} that given $\bmu_i,\;\bSigma_i$
		\begin{eqnarray}
		\bZ_{ij} \mid \bZ_{i(-j)} & \sim & N\left(\bm \theta_{ij},\; \bSigma_{ij}\right)\mathbb{I}\left(\mathbf{Z}_{ij} \in \bm{\Omega}_{ij}\right),
		\end{eqnarray}
		where
		\begin{eqnarray*}
			\bm \theta_{ij} &=& \bm\mu_{ij} + c_{ij}\bm 1_{n_0},\\
			\bSigma_{ij} &=& \bI_{n_0} + \tau_1\bJ_{n_0} + \tau_2(1-f_{2i})\bJ_{n_0},\\
			c_{ij} &=& f_{2i}\left(\bar{Z}_{i(-j)}- \bar{\mu}_{i(-j)}\right),\\
			f_{2i} &=& \frac{n_0(n_{1i}-1)\tau_2}{1+n_0\tau_1+n_0(n_{1i}-1)\tau_2}.
		\end{eqnarray*}
		The $\bm{\Omega}_{ij}$ represents the set of latent variables values which depends on the observed event times as defined in Equation \eqref{omegadef}. 
		
		Second, again using Theorem \ref{thmZsamp}, the conditional distribution of each latent variable $Z_{ijk}$ of patient $i$ given the other latent variables $\bZ_{i(-jk)}$ is derived. Let $\tilde{\tau}_{1i} = \tau_1 + \tau_2(1-f_{2i})$, it then follows that
		\begin{eqnarray}
		Z_{ijk}\mid \bZ_{i(-jk)} & \sim & N(\tilde{\theta}_{ijk},\sigma^2_{ijk})\mathbb{I}\left(Z_{ijk} \in \Omega_{ijk}\right), \label{samplingstepZ}
		\end{eqnarray}
		where
		\begin{eqnarray*}
			\tilde{\theta}_{ijk} &=& \theta_{ijk} + \tilde{c}_{ijk}, \\
			\sigma^2_{ijk} &=& 1 + \tilde{\tau}_{1i}(1-f_{1i}) ,\\
			\tilde{c}_{ijk} &=& f_{1i}\left(\bar{Z}_{ij(-k)}- \bar{ \theta}_{ij(-k)}\right), \\
			f_{1i} &=& \frac{(n_0-1)\tilde{\tau}_{1i}}{1+(n_0-1)\tilde{\tau}_{1i}}.
		\end{eqnarray*}
		This procedure supports the sampling of each latent variable from the joint multivariate distribution, by sequentially sampling from truncated univariate normal distributions.
		
	\subsubsection*{Sampling Covariance Parameters~\label{covariancesampling}}
	The latent variables are re-scaled to have mean zero, $\bV_i = \bZ_i-\bh_i({\bs}_i)-\bX_i\bm\beta$. Under the prior specification stated in Theorem \ref{posteriorstau}, the parameter $\tau_1$ has a shifted inverse-gamma posterior distribution given $\bV_i$:
		\begin{eqnarray}
		\tau_1 \mid \bV_i & \sim & \text{shifted-}\mathcal{IG} \left(\alpha_{\tau_1}+(n_{1i}-1)/2,\; \beta_{\tau_1}+S_{i1}^2/2,\; 1/n_0\right)	\nonumber
		\end{eqnarray}
		where $S_{i1}^2 = \|\bM_{i1}\bH^{(i)}\bV_i\|^2/n_0$, $\bM_{i1} = (\bI_{n_{1i}}\otimes \bK_{n_0})-\bK_{n_0n_{1i}}$ and $\bH^{(i)} = \bH_{n_{1i}}\otimes \bH_{n_0}$. The event times are independently distributed across treatment groups $i$. Thus, the posterior distribution of $\tau_1$ given 
		$\bV$ is given by
		\begin{eqnarray}
		\tau_1 \mid \bV & \sim & \text{shifted-}\mathcal{IG}\left(\alpha_{\tau_1}+\frac{\sum_{i=1}^{n_2}(n_{1i}-1)}{2},\; \beta_{\tau_1}+\frac{\sum_{i=1}^{n_2}
			S_{i1}^2}{2},\; 1/n_0\right).\label{posteriortau1}
		\end{eqnarray}
		
		The design is unbalanced, since the treatment groups have a different number of patients. Hence, the balanced covariance matrix $\bSigma_b$ from \eqref{balancedversion} is considered, for which each covariance matrix $\bSigma_i$ is a principal submatrix: 
		\begin{eqnarray}
		\bSigma_b &=& \bI_{n_0\bar{n}_1} + \tau_1 (\bI_{\bar{n}_1}\otimes \bJ_{n_0}) + \tau_2\bJ_{n_0\bar{n}_1}, \nonumber
		\end{eqnarray}
		where $\bar{n}_1 = \max_i n_{1i}$. From Theorem \ref{constraintbalanced}, it follows that $\bSigma_b$ is positive definite if-and-only-if \eqref{rhoconstraint} and \eqref{tauconstraint} hold, hence the parameter range for $\bm \tau$ is the same under $\bSigma_b$ and each $\bSigma_i$. The $\bV_i$ are augmented with $\bU_i$ to create a vector of observations $\bV_i^b$ which is normally distributed with mean zero and the balanced covariance matrix $\bSigma_b$. Following Theorem \ref{posteriorstau}, the posterior distribution of $\tau_2$ given $\tau_1$ and $\bV^b$ is given by
		\begin{eqnarray}
		\tau_2 \mid (\bV^b,\tau_1) & \sim & \text{shifted-}\mathcal{IG}\left(\alpha_{\tau_2}+n_2/2,\; \beta_{\tau_2}+\frac{\sum_{i=1}^{n_2}S_{i2}^2}{2},\; (1+n_0\tau_1)/(n_0\bar{n}_1)\right)\label{posttau2},\;\;\;\;\;\;
		\end{eqnarray}
		where $S_{i2}^2 = \|\bM_2\bH\bV_i^b\|^2/(n_0\bar{n}_1)$, $\bM_{2} = \bK_{n_0\bar{n}_1}$, and $\bH = \bH_{\bar{n}_1}\otimes \bH_{n_0}.$\\\\ The structure of $\bH$ and $\bM_2$ reveals that the sum of squares $S_{i2}^2$ equals the squared average outcome in each treatment group:
		\begin{eqnarray} 
		S_{i2}^2 & = & \left(\bar{V}_i^b\right)^2 = \left(\frac{n_{1i}\bar{V}_i + (\bar{n}_1-n_{1i})\bar{U}_i}{\bar{n}_1}\right)^2.
		\label{totalmeanformula}
		\end{eqnarray} 
		Thus, the group means $\bar{V}_i$ can be augmented with group means $\bar{U}_i$ to obtain the balanced group means $\bar{V}_i^b$. The Gibbs sampler now makes use of the fact that the conditional distribution of $\bar{U}_i$ given $\bar{V}_i$ and $\bm \tau$ is normal with respective mean and variance 
		\label{samplingUbar}
		\begin{eqnarray}
		E\left(\bar{U}_i \mid \bar{V}_i, \bm\tau \right) &=& \frac{n_0n_{1i}\tau_2}{1+n_0\tau_1+n_0n_{1i}\tau_2}\bar{V}_i ,\nonumber \\
		V\left(\bar{U}_i \mid \bar{V}_i, \bm\tau\right) & = & \frac{1+n_0\tau_1+n_0(\bar{n}_1-n_{1i})\tau_2}{n_0(\bar{n}_1-n_{1i})} - \frac{n_0n_{1i}\tau_2^2}{1+n_0\tau_1+n_0n_{1i}\tau_2}.\nonumber
		\end{eqnarray}
		The fact that $\tau_2$ and $\bar{U}_i$ are sampled from their respective full conditionals given the complete model with balanced covariance matrix $\bSigma_b$ makes that this augmentation procedure is valid (see page \pageref{valid_DA}).

	\section{Simulation Studies\label{simulationstudies}}
	The performance of our Bayesian inference method is evaluated on the data described in Section \ref{datadescr} using two simulation studies. In the first study, the number of treatment groups is large ($n_2=100$), the group sizes $n_{1i}$ are small ($n_{1i}\leq 10$) and five events are recorded per patient ($n_0=5$). The second study mimics the real-life data application introduced in the next section, where three treatment groups are considered which consisted of $1172,\;1169,\text{ and }1173$ patients, for which three events were recorded. Hence, in the second study, the number of treatment groups $n_2$ is set to $3$, $n_{1i}\geq 1169$ and $n_0=3$. For both studies, parameter recovery and coverage rate results are given in Appendix \ref{resultssimstudies}.
	Furthermore, for both studies, plots are shown that display the distribution of the parameter estimates (posterior mean and median) and effective sample sizes over all simulations.
	
	For the first study, on average the parameters are recovered quite well, and the coverage statements expressed by the posterior distribution are close to the actual coverage values, averaged over simulations. This also holds for the second simulation study, but to a lesser extend for the results for $\tau_2$.
	Due to the small number of treatment groups, the posterior variance of $\tau_2$ is very large (small) when $\tau_2$ is far from (close to) zero. As a result, the coverage rate of a $95\%$ coverage interval is larger than expected ($100\%$) when $\tau_2 = 0$, while it is smaller than expected $(87\%)$ for $\tau_2 = 0.025.$ It is concluded that in this situation, while being uncertain about its true value, the algorithm can still reliably determine whether or not $\tau_2\approx 0$.
	
	\section{Real-life Data Application \label{application}}
	Interval-censored clustered event time data from a three-armed randomized clinical trial was analyzed, where patients required a drug-eluting stent during a coronary intervention \cite{von2016very}. The experimental arms in this trial consisted of patients obtaining a biodegradable polymer stent eluting either everolimus ($n_{11}=1172$ patients) or sirolimus ($n_{12}=1169$ patients). The third arm obtained a durable polymer zotarolimus-eluting stent ($n_{13}=1173$ patients). The outcome variables of interest were the event times of  cardiac death, target vessel related myocardial infarction and clinically indicated target vessel revascularization after 12 months (360 days) of follow up. In \cite{von2016very}, log-rank tests were conducted to examine differences in outcomes under the zotarolimus-eluting stent and the everolimus-eluting stent, and between the zotarolimus-eluting stent and the sirolimus-eluting stent separately. They used a primary endpoint, a composite of the three event times, thereby avoiding multiple testing due to dealing with multiple outcome types. However, non-inferiority of the everolimus- and the sirolimus-eluting stents versus the zotarolimus-eluting stent were tested simultaneously. No significant differences were found between the stents (using a significance level of 0.05) with respect to the primary endpoint under the log-rank test. Adjustments for covariates was not necessary, since randomization was seen to have been achieved.    
	
	The registered event times were measured in days for 12 months (360 days). Therefore, right endpoints for survival were set at the registered event times and the left endpoints were set at the right endpoints minus one day. Right-censoring occurred when patients withdrew their consent to be monitored in the study, or when they were lost to follow-up (which occurred rarely, $28/3514\approx 0.80\%$). 
	After cardiac death, patients were not at risk for any of the other events, which violates the non-informative censoring assumption. Therefore, events not observed before cardiac death were both left and right censored and excluded in the sampling of the baseline and covariate effects. This solution, applied to $27/3514\approx 0.77\%$ of the patients, ensured that these patients were retrospectively treated as not being at risk for any of the censored events after occurrence of cardiac death. 
	
	The (semi-parametric) BCSM (survival) model with a two-way nested covariance structure (Equation \eqref{LTMmod}) was used to examine treatment differences. The covariance parameter $\tau_1$ represented the covariance between the observed times of the $n_0=3$ event types of each patient. The covariance between event times of patients in the same treatment group was represented by $\tau_2$. The degree of the monotone splines was set to four, and the spline knots were set at the $20$ equidistant values in the interval $[0.01, 362.6]$. The Gibbs sampler was run with a burn-in of 3,000 iterations. Gibbs sampler convergence was assessed with the Geweke diagnostic (sample size of 500). After convergence was confirmed, the algorithm was applied until an effective posterior sample size of 600 was collected for each non-spline parameter. Improper priors were used for all parameters except for the intra-treatment covariance (see Appendix \ref{resultssimstudies}). A shifted-inverse gamma prior was specified for $\tau_2$ with shape and scale parameter $a_{\tau_2} = 0.001,b_{\tau_2}=0.001$, respectively, which avoided the Gibbs sampler to visit very high values for $\tau_2$. 
	
\begin{figure}[h!]
			\centering
		\includegraphics[width = 0.87\textwidth]{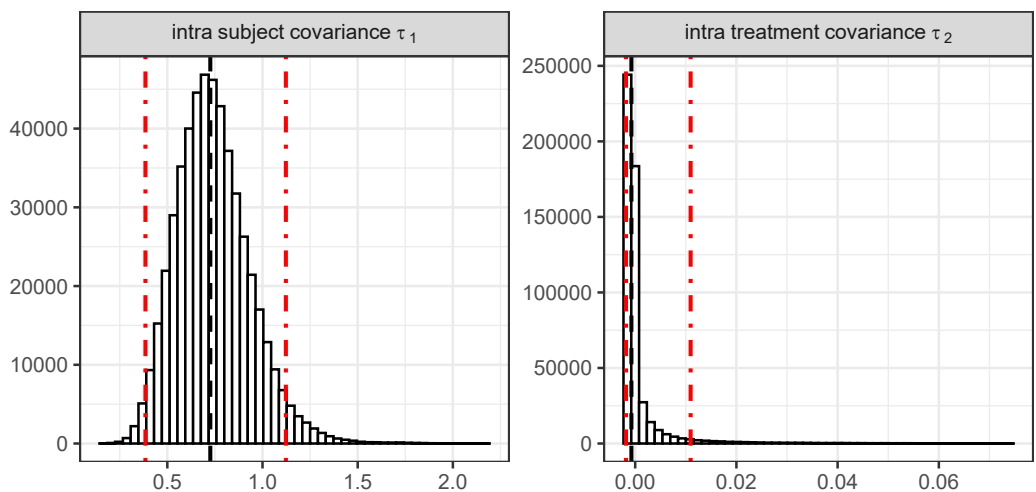}
		\caption{Histogram of posterior samples of $\tau_1$ (left panel) and $\tau_2$ (right panel) under the event time observations in the real-life (BIO-RESORT) dataset. The dashed line shows the median of $\tau_1=0.73$ (left panel) and of $\tau_2=-0.00075$ (right panel). The $95\%$ HPD intervals are represented by the dashed lines with dots.}
		\label{histpoststent}
	\end{figure}

\begin{figure}[h!]
			\centering
		\includegraphics[width  = 0.75\textwidth]{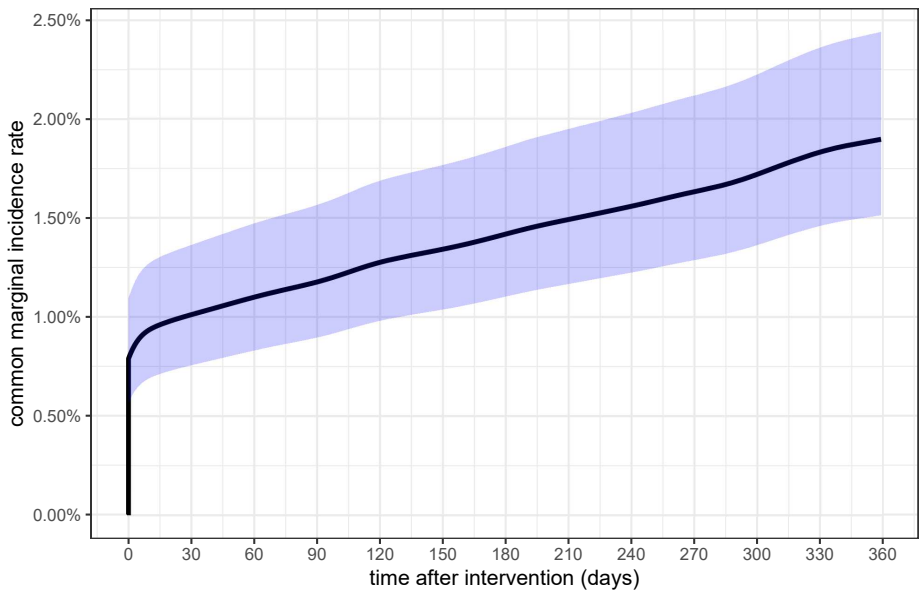}
		\caption{Posterior common marginal incidence rate (cumulative distribution) function for the event time observations in the real-life (BIO-RESORT) dataset. The point-wise mean is plotted vs. the time after intervention, while the shaded part represents a $98\%$ posterior confidence interval.}
		\label{CDFstent}
	\end{figure}

	In Figure \ref{histpoststent}, a histogram is shown of the sampled covariance parameters $\bm{\tau}$. The median and 95\% HPD intervals are also shown. The median intra-patient covariance $\tau_1$ is around $0.73$, and the estimated median correlation is around $0.42$ with a 95\% HPD interval of $[0.29 ,0.54]$. Thus, around $42\%$ of the total variance in observed event times is explained by this moderate correlation among patient's event times. Patients having an event are more likely to receive another event than those not having an event after 12 months follow-up.    
	In Figure \ref{CDFstent}, the median posterior marginal incidence rate (univariate cdf) for any of the considered events is plotted, along with a point-wise $98\%$ credible interval. The (univariate) incidence rate is defined as the probability $P\left(T \le t \right) = \Phi\left(h(t)/(1+\tau_1+\tau_2)\right)$, using a common baseline for the different event types. The plot shows the marginal probability of any event before each day within the 12 months of follow-up. The marginal incidence rate after one year was very low and one of the challenging aspects. It can be seen that the credible interval lies close to the median incidence rate. Furthermore, the median posterior incidence rate is in agreement with the reported trend in composite incidence rates of cardiac death, myocardial infarction and target vessel revascularization, which was computed with the Kaplan-Meier estimator (see \cite{von2016very} Fig. 2A).
	
	\subsection{Treatment Differences}
	Substantial evidence in favor of the hypothesis $\tau_2 \le 0$ corresponded to equivalence between the treatment arms (i.e., the three stent types performed similar). The hypothesis $\tau_2>0$ represented a difference in performance of the three treatment arms (non-equivalence). The posterior distribution of $\tau_2$ included the hypothesis of negative, positive and no within-cluster dependence. Therefore, highest posterior density (HPD) interval and posterior odds testing were used to examine treatment differences for three treatment arms on three different endpoints. Our approach avoided testing multiple null hypotheses by collecting data evidence from each treatment arm on each endpoint in favor of the null hypothesis of equivalence (i.e., no risk differences) -- otherwise this would be a multiple testing problem --. Furthermore, composite endpoints were not required, which can lead to incorrect statistical inferences, and the individual component endpoints can be expected to provide more data evidence about treatment effects.     
	
	The estimated median intra-treatment covariance is around $-0.00075$ with a $95\%$ HPD interval of $[-0.0018, 0.011]$. The point zero is included, and it can be concluded that there is no clustering effect by treatment groups (i.e., equivalence of stent groups). The posterior probability of a positive covariance ($P(\tau_2>0 \mid \mathbf{L},\mathbf{R})$) is around $0.23$, which shows the weak support in favor of treatment differences. The posterior probability of a non-positive correlation, $P(\tau_2 \le 0 \mid \mathbf{L},\mathbf{R})$ is around $0.77$. Thus, there is more evidence for equal risk of the three stents with respect to the three event types. 
	
	It follows that the Bayes factor of \textit{no risk differences} ($\tau_2\leq 0$) against \textit{any risk differences} $(\tau_2>0)$ was around $0.77/0.23\approx 3.35$, under equal prior probability for the two hypotheses. This result can be classified as substantial evidence in favor of equivalence of risk for the three stents. 
	
	A log-rank (one-sided) P-value of $0.23$ was computed for testing the non-inferiority hypothesis (with margin $0$) of the sirolimus-eluting stent versus the zotarolimus-eluting stent on the composite endpoint. The same P-value was obtained for non-inferiority of the everolimus-eluting stent versus the zotarolimus-eluting stent. Following \cite{marsman2017three}, it was expected that the posterior probability of superiority of the zotarolimus-eluting stent over another stent was equal to the one-sided P-value of the standard log-rank test. Then, under equal prior probability for the two hypotheses,
	the corresponding Bayes factor in favor of non-inferiority of the sirolimus-eluting stent was around $0.77/0.23\approx 3.35$, and showed substantial evidence in favor of non-inferiority. The same result was obtained for testing non-inferiority of the everolimus-eluting stent. Interestingly, this result was similar to the Bayes factor result of equivalence of risk differences. However, note that the Bayes factor for non-inferiority represented a single test for the non-inferiority of a specific stent, where the Bayes factor for equivalence of risk examined if at least one of the three stents had a different risk. 
	
	Furthermore, standard (two-sample) log-rank tests were performed to assess the null hypothesis of no difference in survival between the sirolimus- versus zotarolimus-eluting  (P-value $0.46$) and everolimus- versus zotarolimus-eluting (P-value $0.45$) stents on the composite endpoint \cite{von2016very}. By representing the point null hypothesis as a combination of two one-sided tests  \cite{shi2020reconnecting}, the P-value for the two-sided hypothesis test was expected to be equal to two times the posterior probability of superiority of the zotarolimus-eluting stent, and subsequently was also around $0.23$. Thus, similar to the one-sided log-rank tests, data evidence was found in favor of no difference. 
	
	The median lower bound for $\tau_2$, induced by the positive definiteness requirement of the covariance matrix, can be estimated using posterior samples of $\tau_1$ using \eqref{tauconstraint}. This yielded an estimated lower bound of $-0.00090$ and shows that the posterior distribution of $\tau_2$ was concentrated around this lower bound (see right-panel of Figure \ref{histpoststent}).  A negative correlation indicates that patients who received the same treatment showed (unobserved) heterogeneity in their event times. Unobserved factors such as the heart surgeon or the medical centre could explain heterogeneity between patients receiving the same treatment. Furthermore, within-treatment heterogeneity could relate to differences in patient-related factors such as smoking, diabetes, and stent length. These factors did not affect between-treatment differences due to the randomization of patients to treatments. However, ignoring a negative cluster correlation still leads to a deflation of the Type-1 error and conservative behavior of a test for intra-treatment differences.  
	
	\section{Discussion \label{discussion}}
	A BCSM modeling framework for multi-way nested data is proposed. It includes a model for the structured covariance matrix to describe associations among clustered observations. The structured covariance model can be integrated in a multivariate linear transformation model, which enables a joint analysis of multivariate event times with a nested classification structure. Directly modeling the dependence structure through a structured covariance matrix has several advantages. When adding another nested factor to the model, the complexity of the model increases with only one covariance parameter. This makes the model particularly useful for higher-order nested structures in which multiple layers of dependencies are defined. Although clustered event times are often positively correlated, they can also be negatively correlated (for example, see \cite{xue1996bivariate}). With the structured covariance matrix negatively and positively correlated observations can be modelled. Furthermore, in the real-life data application it sufficed to assume a common baseline survival function, but it is also possible to have different baselines across event types.     
	
	The use of random effect parameters for unobserved heterogeneity in the population has several disadvantages. The inclusion of random effect parameters increases the model complexity rapidly, and leads to multidimensional integrals in the likelihood. The random effect parameters induce sample size restrictions, since between and within-cluster variance components need to be estimated. The estimation of population-average regression effects is complicated, since regression coefficients in the model are defined conditionally on the random effect parameters. Furthermore, shared random effects will only induce positive correlation among clustered event times, which makes them not suitable for modeling negatively correlated event times.  
	
	\subsection{Testing the Structured Covariance Matrix}
	In a marginal modeling approach \cite{prentice2019statistical}, marginal distributions of multivariate event times are formulated without specifying the nature of dependence among clustered event times. In our BCSM modeling approach, associations modeled by the structured covariance matrix are of specific interest. Associations between clustered event times at different levels are described by covariance parameters. Hypotheses concerning restrictions on the covariance parameters are of interest to examine for instance differences between treatment effects or in risks between groups. The flexible shifted-inverse gamma prior for the covariance parameter supports testing hypotheses concerning negative, positive, or no intra-cluster dependencies. 
	
	A positive covariance parameter represents between-cluster differences. Therefore, the problem of testing equivalence of multiple cluster means can be reformulated to testing whether a single covariance parameter is zero. This approach transforms the multiple testing problem (i.e., $\mu_i \ne \mu_j$) to a single testing problem ($\tau_q\neq0$). For multiple treatment groups, the standard approach is to compare all groups against each other to assess the treatment effect. However, this method has the well-known disadvantage that the P-value is inflated, and correction methods for a family of hypotheses are complex and not optimal. Difficulties associated with multiple comparison procedures only increase in the number of subgroups to analyze. Furthermore, for group-mean comparisons the statistical power is lowered due to a smaller available sample size, while additional heterogeneity in the subgroup outcomes is often ignored. The BCSM model can be extended to include non-nested (cross-classified) dependencies. An interesting application is to model the covariance among event times of the same type, and to test for heterogeneity in risk across event types. This test could be used to examine the support for event type specific baseline hazards.  
	
	\subsection{Multiple Endpoints}
	An important treatment effect cannot always be identified by evaluating a single endpoint, in particular when the event type occurs with a low frequency. A composite endpoint can then be constructed from component endpoints to increase the number of events and to achieve adequate statistical power for a study. A main advantage is that a single hypothesis can be evaluated to show superiority of the treatment on the composite endpoint, where simultaneous tests are required to show superiority on all endpoints. In our multivariate modeling approach, treatment effects can be examined for multiple endpoints in a more straightforward manner by evaluating a single hypothesis, while allowing a heterogeneous relationship between event times from the same subjects.  
	
	The BCSM model for multiple endpoints can be generalized by allowing treatment differences across endpoints, which corresponds to modeling event type specific covariance parameters and to test for a treatment effect for each endpoint. Following the approach of \cite{gonen2003bayesian}, who evaluate the equality of means between regimens, a prior on the hypotheses can be included to address their correlation.      
	
	Under the derived Bayesian inference procedure, it is possible to conduct Bayesian covariance tests using Bayes factors \cite{Fox2010bayesian,mulder2019bayes}. Bayes factors have some advantages over frequentist hypothesis tests \cite{wagenmakers2018bayesian}. First, the compared hypotheses are clearly defined. Next, in Bayes factor testing no pre-specified significance level has to be set, as compared to frequentist tests. Instead, one can just report the observed evidence for each of the hypotheses. Data collection procedures can be altered intermediately due to the fact that Bayesian inference is insensitive to the data collection procedure.  

	

\section{Acknowledgements}
The authors thank Clemens von Birgelen, Carine Doggen and Eline Ploumen for their collaboration and the use of their data for the real-life data application.

\bibliographystyle{unsrt}
\bibliography{BCSM_multi_way_nested_data_Baas_Boucherie_Fox}

\newpage
\appendix

	\section{Conditional Distribution of Latent Variables}\label{prooftheorem32}
	The proof of Theorem 4 is given. 
	\begin{proof}
		Without loss of generality, the covariance matrix is partitioned
		\begin{eqnarray*}
			\bSigma &=& \begin{bmatrix}
				\bSigma^{(j)}&\bSigma^{(j(-j))} \\
				\bSigma^{((-j)j)}&\bSigma^{(-j)}
			\end{bmatrix}
		\end{eqnarray*}
		where $\bSigma^{(j)}$ is the covariance matrix of $\mathbf{Z}_{ij}$. Using a well-known property of the normal distribution, the conditional distribution of $\mathbf{Z}_{ij}$ given $\mathbf{Z}_{i(-j)}$ is multivariate normal with mean and covariance:
		\begin{align*}
		&\bm\theta_{ij} = \bm\mu_{ij} +  \bSigma^{(j(-j))}\left[\bSigma^{(-j)}\right]^{-1}\left(\mathbf{Z}_{i(-j)}-\bm    \mu_{i(-j)}\right) \\
		&\bSigma_{Q-1}   = \bSigma^{(j)}  - \bSigma^{(j(-j))}\left[\bSigma^{(-j)}\right]^{-1}\bSigma^{((-j)j)},
		\end{align*}
		respectively. The expressions are rewritten to obtain the specified mean and covariance.\\
		Using Lemma 3 on $\bSigma^{(-j)}$, it holds that
		\begin{align*} 
		\bSigma^{(j(-j))} &= \tau_Q(\bm 1^\top_{m_{Q-1}-1}\otimes \bJ_{s_{Q-1}}) \\
		\left[\bSigma^{(-j)}\right]^{-1}&= \frac{1}{\tau_0}\bI_{u_{Q}} + \sum_{q=1}^{Q-1} \rho_q (\bI_{\tilde{m}_q}\otimes \bJ_{s_q}) - \frac{f_Q}{v_{Q-1}u_Q} \bJ_{u_Q},
		\end{align*}
		where $\tilde{m}_q = m_q(n_{Q-1}-1)/n_{Q-1}$ and $\rho_q$ are defined in (12). Hence,
		\begin{eqnarray*}	\bSigma^{j(-j)} \left[\bSigma^{(-j)}\right]^{-1} &=& \frac{\tau_Q}{\tau_0}(\bm 1_{m_{Q-1}-1}^\top\otimes \bJ_{s_{Q-1}}) +  \tau_Q\sum_{q=1}^{Q} \rho_q (\bm 1_{m_{Q-1}-1}^\top\otimes \bJ_{s_{Q-1}})(\bI_{\tilde{m}_q}\otimes \bJ_{s_q}) \\ 
			&& -\frac{f_Q\tau_Q}{v_{Q-1}u_Q} (\bm 1_{m_{Q-1}-1}^\top\otimes \bJ_{s_{Q-1}})\bJ_{u_Q} \\
			&=& \frac{\tau_Q}{\tau_0}(\bm 1_{m_{Q-1}-1}^\top\otimes \bJ_{s_{Q-1}}) +  \tau_Q\sum_{q=1}^{Q} \rho_q (\bm 1_{m_{Q-1}-1}^\top\otimes \bJ_{s_{Q-1}/s_q}\otimes \bJ_{s_q})(\bI_{\tilde{m}_q}\otimes \bJ_{s_q}) \\
			&& -\frac{f_Q\tau_Q}{v_{Q-1}} (\bm 1_{m_{Q-1}-1}^\top\otimes \bJ_{s_{Q-1}})\\
			&=&\frac{\tau_Q}{\tau_0}(\bm 1_{m_{Q-1}-1}^\top\otimes \bJ_{s_{Q-1}}) +  \tau_Q\sum_{q=1}^{Q-1}s_q \rho_q  (\bm 1_{m_{Q-1}-1}^\top\otimes \bJ_{s_{Q-1}}) \\
			&& - \frac{f_Q\tau_Q}{v_{Q-1}}(\bm 1_{m_{Q-1}-1}^\top\otimes \bJ_{s_{Q-1}}) \\
			&=& \tau_Q\left(\frac{1}{\tau_0}+\sum_{q=1}^{Q-1} s_q\rho_q - \frac{f_Q}{v_{Q-1}} \right)(\bm 1_{m_{Q-1}-1}^\top\otimes \bJ_{s_{Q-1}}) = \frac{f_Q}{u_Q}(\bm 1_{m_{Q-1}-1}^\top\otimes \bJ_{s_{Q-1}}). 
		\end{eqnarray*} 
		The last equation follows by plugging in $1/v_{Q-1} = 1/\tau_0+\sum_{q=1}^{Q-1} s_q\rho_q$ and using that $u_Q\tau_Q(1-f_Q)=v_{Q-1}f_Q$. It follows that the conditional mean is given by
		\begin{eqnarray*}
			\bm\theta_{ij} &=& \bm\mu_{ij} +  \frac{f_Q}{u_Q}(\bm 1_{m_{Q-1}-1}^\top\otimes \bJ_{s_{Q-1}})(\bZ_{i(-j)}-\bm \mu_{(i(-j))}) = \bm\mu_{ij} +c_{ij}\bm 1_{s_{Q-1}},\nonumber 
		\end{eqnarray*}
		and the conditional covariance matrix can be obtained from,
		\begin{eqnarray*}
			\bSigma^{(j(-j))}\left[\bSigma^{(-j)}\right]^{-1}\bSigma^{((-j)j)} &=& \tau_Qf_Q\bJ_{s_{Q-1}}\\
			\implies \bSigma_{Q-1} &=& \tau_0\bI_{s_{Q-1}} +  \sum_{q=1}^{Q-1} \tau_q (\bI_{m'_q}\otimes \bJ_{s_q}) + \tau_{Q}\left(1-f_Q\right)\bJ_{s_{Q-1}}.
		\end{eqnarray*}
	\end{proof}
	
	\section{Conditional Distribution of Covariance Parameters}\label{prooftheorem33}
	The proof of Theorem 5 is given. 
	\begin{proof}
		To derive the result, the covariance matrix of $\bW_{iq}:=\bM_q\bH\bV_i$ is derived. Therefore, consider the covariance matrix of each $\bH\bV_i$
		\begin{align*}
		\bH\bSigma\bH^\top &= \tau_0\bH\bH^\top + \sum_{q=1}^Q\tau_q\left(\bigotimes_{r={Q-1}}^0 \bH_{n_r}\right)\left(\left(\bigotimes_{r= Q-1}^{q}\bI_{n_{r}}\right)\otimes \left(\bigotimes_{r = q-1}^{0} \bJ_{n_{r}}\right)\right)\left(\bigotimes_{r=Q-1}^0 \bH_{n_r}^\top\right)\\&=\tau_0\bI_{s_Q} + \sum_{q=1}^Q\tau_q\left(\bigotimes_{r= Q-1}^{q}\bH_{n_{r}}\bH_{n_{r}}^\top\right)\otimes \left(\bigotimes_{r = q-1}^{0} \bH_{n_{r}}\bJ_{n_{r}}\bH_{n_{r}}^\top\right)\\&= \tau_0\bI_{s_Q} + \sum_{q=1}^Q\tau_q\bI_{m_q}\otimes \left(\bigotimes_{r = q-1}^{0}n_{r}\bK_{n_{r}}\right) = \tau_0\bI_{s_Q} + \sum_{q=1}^Q\tau_qs_q\left(\bI_{m_q}\otimes\bK_{s_{q}}\right).
		\end{align*}
		For all $r,q$, it holds that
		\begin{eqnarray*}
			\left(\bI_{m_q}\otimes \bK_{s_q}\right)\left(\bI_{m_r}\otimes \bK_{s_r}\right) = \bI_{m_q\wedge m_r}\otimes \bK_{s_q\vee s_r},
		\end{eqnarray*}
		with $m_q\wedge m_r\equiv\min(m_q,m_r)$ and $s_q\vee s_r\equiv\max(s_q,s_r)$. From this, it follows that
		$$\bM_q(\bI_{m_r}\otimes \bK_{s_r})\bM_q^\top = \left(\bm \bI_{m_{q}\wedge m_r}\otimes  \bK_{s_{q}\vee s_r}\right)-
		\left(\bI_{m_{q+1}\wedge m_{r}}\otimes \bK_{s_{q+1}\vee s_{r}}\right).$$
		Hence, for $q\in\{0,1,\dots, Q-1\}$, letting the sum above run over $r$ instead of $q$, the covariance matrix of $\bW_{iq}$ is equal to
		\begin{align}
		\bM_q\bH\bSigma\bH^\top\bM_q^\top &= \tau_0\bM_q + \sum_{r=1}^Q\tau_rs_r\left(\left(\bm \bI_{m_{q}\wedge m_r}\otimes  \bK_{s_{q}\vee s_r}\right)-
		\left(\bI_{m_{q+1}\wedge m_{r}}\otimes \bK_{s_{q+1}\vee s_{r}}\right)\right)\nonumber\\&= \tau_0\bM_q + \left(\left(\bm \bI_{m_{q}}\otimes  \bK_{s_q}\right)-
		\left(\bI_{m_{q+1}}\otimes \bK_{s_{q+1}}\right)\right)\sum_{r=1}^q\tau_rs_r\nonumber\\&=
		\left(\tau_0+\sum_{r=1}^q\tau_rs_r\right)\bM_q.\label{covmatk}
		\end{align} 
		Similarly
		\begin{align}
		&\bM_Q\bH\bSigma\bH^\top\bM_Q^\top= \left(\tau_0+\sum_{r=1}^Q\tau_rs_r\right)\bM_Q.\label{covmatQ}\end{align}
		The only nonzero entries of the diagonal covariance matrix of $\bW_{iq}$ are hence equal to $$\sigma_q^2 := \tau_0+\sum_{r=1}^q\tau_rs_r.$$ 
		The matrix $\bH$ is invertible, hence the variables $(\bH\bV_i)_{i=1}^{n_Q}$ are sufficient for $\bm \tau$. 
		Next, there is a one-to-one relation between $\bm \sigma^2$ and $\bm \tau$. Hence, $\bW$ is also sufficient for $\bm \sigma^2$. From the prior specification for $\bm \tau$, it can be seen that $\sigma^2_0,\dots,\sigma^2_Q$ are independent a priori with 
		\begin{eqnarray*}
			\sigma^2_q &\sim& \mathcal{IG}(\alpha_{\tau_q},\beta_{\tau_q}). 
		\end{eqnarray*}
		The $p_q=\text{tr}(\bM_q)$ nonzero entries of each $\bW_{iq}$ are normally distributed with mean zero and variance $\sigma^2_q$. it follows that -- given the prior specification and the previous analysis --
		\begin{eqnarray*}
			\sigma_q^2 \mid \bV \stackrel{d}{=}\sigma_q^2|\bW \sim  \mathcal{IG}\left(\alpha_{\tau_q}+n_Qp_q/2,\; \beta_{\tau_q} + \sum_{i=1}^{n_Q}\|\bW_{iq}\|^2/2\right),
		\end{eqnarray*}
		where $\stackrel{d}{=}$ denotes equality in distribution. Finally, the posterior distribution of $\bm \tau$ can be derived by transforming $\bm \sigma^2$ back to $\bm \tau$ in the above posterior distribution.
	\end{proof}

		\section{Sign Preservation of Covariance Under Monotone Transformations}\label{covarianceproof}
	In Section 4 of the main paper, the latent vectors $\bZ_i$ are multivariate normally distributed and an affine function of the transformed event times $\bh_i(\bt_i)$.	From this, it follows that $\bt_i$ is distributed as $\bh_i^{(-1)}$ applied to an affine function of $\bZ_i$, where $\bh^{(-1)}_i$ is the nondecreasing generalized inverse of $\bh_i$ defined as (3.1) in \cite{de2015study}.
	
	The statistical inference procedure is designed to estimate the covariance parameters $\btau$ of the covariance matrix $\bSigma_i$ of $\bZ_i$. Intu\"itively, it makes sense that due to the nondecreasing nature of $\bh_i^{(-1)}$, the covariance between event times, $\bt_i$, has the same sign as the covariance (positive, zero or negative) between the respective latent variables $\bZ_i$, conditional on mean differences through explanatory variables. Vice versa, if the event times show a different behavior than expected from the sign of the covariance of the latent variables, this can only be due to perturbation of this behavior by the effects of explanatory variables. The following theorem shows that this consistency property of the sign of the covariance indeed holds.
    Furthermore, due to exchangeability, clusters of latent variables that are equicorrelated correspond to equicorrelated event times. 
		\begin{theorem} \label{theo}
		Let $[X,Y]^\top\in\mathbb{R}^2$ be a multivariate normally distributed vector, i.e. for some $\mu_x,\mu_y, \rho_{xy}\in\mathbb{R}$ and $\sigma_x,\sigma_y \in (0,\infty)$ such that $|\rho_{xy}| \leq \sigma_x\sigma_y $: $$\begin{bmatrix}
		X\\Y\end{bmatrix}\sim N\left(\begin{bmatrix}\mu_x\\\mu_y\end{bmatrix},\; \begin{bmatrix}\sigma_x^2&\rho_{xy}\\\rho_{xy}&\sigma_y^2\end{bmatrix}\right)
		.$$ 
		Let $f$ and $g$ be nondecreasing, non-constant functions such that the random variables $f(X)$ and $g(Y)$ have a finite second moment. Then, it holds that
		\begin{eqnarray}
		    \text{sgn}\left(\text{Cov}\left(f(X),g(Y)\right)\right) &=& \text{sgn}(\rho_{xy}).\nonumber
        \end{eqnarray}
	\end{theorem}
	\begin{proof}
		For a random variable $Z$ and two non-decreasing functions $\phi,\psi$ such that random variables $\phi(Z),\psi(Z)$ have a finite second moment, it holds that $\text{Cov}\left(\phi(Z),\psi(Z)\right)\geq 0$ \cite{schmidt2014inequalities}.	From the proof in \cite{schmidt2014inequalities}, it can be seen that $\text{Cov}\left(\phi(Z),\psi(Z)\right)= 0$ if and only if either the random variable $\psi(Z)$ or the random variable $\phi(Z)$ is deterministic. The if-part follows directly from the Cauchy-Schwarz inequality and heuristically the only-if part follows from the fact that $\phi(Z)$ has to ``increase together" with $\psi(Z)$ if both random variables are able to attain multiple values.
		
		\noindent Now, by the law of total expectation:
		\begin{eqnarray}
		\text{Cov}(f(X), g(Y)) &=& E[f(X)g(Y)] - E[f(X)] \cdot E[g(Y)] \nonumber \\
		&=& E[f(X)E\left[g(Y)|X\right]] - E\left[f(X)\right]\cdot E\left[E\left[g(Y)|X\right]\right].
		\nonumber 
		\end{eqnarray}
		As $E[g(Y)|X]$ is a function of $X$, say $\tilde{g}(X)$, it holds by the above that: 
		\begin{equation}
		\text{Cov}(f(X),g(Y))=\text{Cov}(f(X), \tilde{g}(X)).
		\label{omgeschreven}
		\end{equation}
		As $g(Y)$ has a finite second moment, $\tilde{g}(X)$ has a finite second moment too due to Jensen's inequality for conditional expectation. By the well-known formula for conditional normal distributions, it now holds that 
		\begin{eqnarray*}
			Y|X\sim N(m(X), \sigma^2)
		\end{eqnarray*}
		where $m(X) = \mu_y + \frac{\rho_{xy}}{\sigma_x^2}(X-\mu_x)$, and $\sigma  = \sqrt{\sigma_y^2 - \frac{\rho_{xy}^2}{\sigma_x^2}}$. Hence,
		\begin{eqnarray}
		\tilde{g}(X)=E\left[g(Y)|X\right] &=& \int_{\mathbb{R}}g(y)\frac{1}{\sqrt{2\pi }\sigma}e^{-\frac{(y-m(X))^2}{2\sigma^2}}dy \nonumber \\
		&=& \int_{\mathbb{R}}g\left( z+m(X)\right)\frac{1}{\sqrt{2\pi}\sigma}e^{-\frac{z^2}{2\sigma^2}}dz \nonumber \\
		&=& E[g(Z+m(X))],\nonumber
		\end{eqnarray}
		where $Z\sim N(0,\sigma^2)$ independent of $X$. It follows that $\tilde{g}(x)$ is nondecreasing and non-constant when $\rho_{xy}>0$, constant when $\rho_{xy} = 0$ and nonincreasing and non-constant when $\rho_{xy}<0.$ This is because the slope of the linear function $m(x)$ has the same sign as $\rho_{xy}$, and $g(Z+y)$ is nondecreasing in $y$ almost surely. For $\rho_{xy}\geq 0$, the statement of the theorem follows from Equation (\ref{omgeschreven}) and the first paragraph of the proof. This is because the support of $X$ is $\mathbb{R}$ and hence the random variable $\tilde{g}(X)$ is deterministic if and only if $g$ is a constant function on $\mathbb{R}$, which we assumed not to be the case. For $\rho_{xy}<0$, one can use the previous results to establish $$\text{Cov}(f(X),g(Y))=\text{Cov}(f(X), \tilde{g}(X)) = -\text{Cov}(f(X), -\tilde{g}(X))<0.$$
	\end{proof}
	\section{Gibbs Sampler \label{Gibbssampler}}
    For the model in Equation (20), the full conditionals for all parameters are combined to form a Gibbs sampler.
    
    We first define some terminology related to the baseline specification, which is modelled as a linear combination of monotone splines, in accordance with \cite{lin2010semiparametric}.
    Integers $K$ and $d$ are introduced to specify monotone spline basis functions of degree $d$ and a knots vector $\bm{\kappa}\in\mathbb{R}^{K+d}$ with $\kappa_1=\kappa_2=\dots=\kappa_d$, $ \kappa_{K+1}=\kappa_{K+2} =\dots=\kappa_{K+d}$ and $\kappa_l< \kappa_{l+1}$ for $l\in\{d,d+1,\dots ,K\}$.
    
    Each monotone spline basis function, denoted as $B_l^{d,\bm{\kappa}}$, is then a $d-1$ continuously differentiable function that has the form of a nonnegative, nondecreasing degree $d$ polynomial on the interval $[\kappa_l, \kappa_{l+d}]$.
    A coefficient vector $\bgamma\in\mathbb{R}^{K+1}$ is defined such that $\gamma_1\in\mathbb{R}$ and $\bm{\gamma}_{(-1)}\in\mathbb{R}_+^{K}$ to specify the baseline as, 
	\begin{equation}
	h_{ijk}(t|\;\bgamma) = \gamma_1 + \sum_{l=2}^{K+1}\gamma_lB_{l-1}^{d,\bm{\kappa}}(t)\;\;\;\; \forall t\in (0,\infty).\label{baselineformula}
	\end{equation}
Note that due to the range of $\bgamma$, the baseline is a non-decreasing function. Let $\bm{\gamma}_{(-\ell)}$ denote the vector of spline coefficients with the $\ell$-th entry removed. 

Let $\bB_{ijk}=[1,B_1^{d,\bkappa}(s_{ijk}),\dots, B_K^{d,\bkappa}(s_{ijk})]$ and let $\bB_i$ be the matrix with rows $\bB_{ijk}$ such that $\bB_i\bgamma = \bh_i(\bs_i)$. Let $\bB$ be the vertical concatenation of $\bB_i$ and $\bb_{\ell}$ be its $\ell$-th column, denote with $\bB_{(-\ell)}$ the matrix $\bB$ with column $\ell$ removed. Let the matrices $\bZ,\;\bX$ be the concatenation of $\bZ_i,\;\bX_i$ (resp) and $\bSigma$ be a block diagonal matrix with blocks $\bSigma_i$ concatenated in the same order as used in the construction of $\bB$. Let $\tilde{\bR}_{ijk}=\left[1,B_1^{d,\bkappa}\left(R_{ijk}\right),\dots, B_K^{d,\bkappa}\left(R_{ijk}\right)\right]$ such that $\tilde{\bR}_{ijk}\bm\gamma= h\left(R_{ijk}\right)$ and let $\tilde{\bL}_{ijk}$ be defined similarly.
	
	Convergence is sped up by initializing the covariate effects $\bbeta$ and spline coefficients $\bgamma$ on their maximum likelihood estimates, where the covariance parameters are fixed to zero. The likelihood of the data under this assumption of ``zero covariance" is given by:
	\begin{equation}
	f(\bbeta,\bgamma):= \prod_{i=1}^{n_2}\prod_{j=1}^{n_{1i}}\prod_{k=1}^{n_0}\left(\Phi\left(\tilde{\bR}_{ijk}\bgamma + \bx_{ijk}^\top\bbeta\right) - \Phi\left(\tilde{\bL}_{ijk}\bgamma + \bx_{ijk}^\top\bbeta\right)\right).\label{likelihoodnocov}
	\end{equation}
	This likelihood is maximized with respect to $(\bbeta,\bgamma)$ with a constrained nonlinear optimization algorithm to initialize the parameters $\bbeta, \bgamma$. 
	
	The Gibbs sampler for sampling $(\bm \beta,\bm\gamma,\bm \tau)$ is given in Algorithm \ref{MCMCalg}. The full conditionals of $\bm \beta,\bm \gamma$ and a hierarchical parameter $\eta$ are described in separate steps. For ease of notation, it is made implicit below that every random variable is sampled according to its full conditional distribution, using the most recent version of the parameters in every iteration of the Gibbs sampler. 
	
	\begin{algorithm}[h!]\setstretch{1.35}
		\caption{Gibbs sampler for posterior inference under the model in (20) under type-II interval censored survival data.}\label{MCMCalg}
		\begin{algorithmic}[1]
			\Inputs{Observed data $(\bL,\bR,\bX)$, number of iterations $M$\\ Prior parameters $\left(\bbeta_0, \bm\Lambda_{0}, v_0,m_0,\alpha_\eta,\beta_\eta,\alpha_{\tau_1},\beta_{\tau_1},\alpha_{\tau_2},\beta_{\tau_2}\right)$\\
				Spline knots $\bkappa$, spline degree $d$;}
			\vspace{2mm}\Initialize{Set $\bbeta^{(0)} = \hat{\bbeta}$ and $\bgamma^{(0)}=\hat{\bgamma}$, where $(\hat{\bbeta},\;\hat{\bgamma})$ are the numerical maximizers of $f$ in (\ref{likelihoodnocov});\\
				Set $\tau_1^{(0)}=0$, $\tau_2^{(0)}=0$, $\eta^{(0)}=1$;\\
				Sample $Z_{ijk} \sim N(\bX_{ijk}^\top\bbeta^{(0)} + \bB_{ijk}\bgamma^{(0)},\; 1)\mathbb{I}(Z_{ijk}\in\Omega_{ijk})$ for all $i,j,k$;}
			\vspace{1mm}\For{$m\in\{1,\dots,M\}$ }
			\For{all $i,j,k$}
			\State Sample $Z_{ijk}$ according to (27);
			\EndFor
			\State Sample $\bbeta^{(m)}$ using $\bm \beta_0$ and $\bLambda_0$ (explained below);
			\For{$\ell \in \{1,\dots, K+1\}$}
			\State Sample $\bgamma^{(m)}_\ell$ using $m_0,v_0,\eta^{(m)}$ (explained below);
			\EndFor
			\State Sample $\eta^{(m)}$ using $\alpha_\eta,\beta_{\eta}$ (explained below);
			\State Sample $\tau_1^{(m)}$ according to (28);
			\State Sample $\bar{\bU}_i$ Gaussian with mean, variance from Subsection 4.1;
			\State Calculate $S_{i2}^2$ using (30);
			\State Sample $\tau_2^{(m)}$ according to (29);
			\EndFor\\
			{\bf Outputs:}\\
			$\;\;\;\;\left(\bbeta^{(1)},\bgamma^{(1)},\tau^{(1)},\rho^{(1)}\right),\dots, \left(\bbeta^{(M)},\bgamma^{(M)},\tau^{(M)}, \rho^{(M)}\right).$
		\end{algorithmic}
	\end{algorithm}
\newpage
	\begin{itemize}
		\item Sampling Covariate Effects $\bm \beta$\\
		From the likelihood (23), it can be seen that when $\mathbf{V}_i := \bSigma_i^{-1/2}\left(\mathbf{Z}_i-\bh_i\left(\bs_i|\;\bgamma\right)\right)$, that $(\mathbf{V},\mathbf{X})$ are sufficient statistics for $\bm{\beta}$. Furthermore, $\mathbf{V}_i$ follows a linear regression model with regression parameters $\bm{\beta}$:
		$$\mathbf{V}_i = \left(\bSigma_i^{-1/2}\mathbf{X}_i\right)\bm{\beta} + \mathbf{E}_i,\;\;\;\; \mathbf{E}_i\sim N\left(\mathbf{0}, \bI_{s_Q}\right).$$
		Assume a $N(\bbeta_0,\mathbf{\Lambda}_{0}^{-1})$ prior on $\bm{\beta}$, where $\mathbf{\Lambda}_{0}$ is a precision matrix. From this prior specification, it now follows that posterior distribution is multivariate normal, and one can sample $\bm{\beta}$ from the full conditional distribution as follows:
		\begin{align}
		&\bm{\beta}|(\bL,\bR,\bX,\bZ,\bm \gamma,\bm \tau)\stackrel{d}{=}\bm{\beta}|(\mathbf{V},\mathbf{X}) \sim N\left(\bmu_{\bbeta},\; \mathbf{\Sigma}_{\bbeta}\right)\label{samplebeta}\\&\text{where}\nonumber\\&\mathbf{\Sigma}_{\bbeta} = \left(\mathbf{\Lambda}_{0} + \sum_i \mathbf{X}_i^\top\mathbf{\Sigma}_i^{-1}\mathbf{X}_i\right)^{-1}\nonumber,\\
		&\bmu_{\bbeta} = \mathbf{\Sigma}_{\bbeta}\left(\mathbf{\Lambda}_{0}\bbeta_0 - \sum_i \mathbf{X}_i^\top\mathbf{\Sigma}_i^{-1}\left(\mathbf{Z}_i-\bh_i\left(\bs_i|\;\bgamma\right)\right)\right).\nonumber
		\end{align}
		
		
		\item Sampling Spline Coefficients $\bm\gamma$\\
		From the likelihood (23), it can be seen that $\gamma_\ell$ can be sampled from the full conditional distribution in a manner similar to the one in Lin and Wang (2010), with a few alterations. 
		Assume that the prior distribution for $\gamma_1$ is a normal distribution with mean $m_0$ and variance $1/v_0$. Furthermore, assume an exponential prior with a hierarchical parameter $\eta$ for all $\gamma_1,\dots, \gamma_K$. 
			To regularize the sampled baseline parameters $\bgamma$, a $\Gamma(\alpha_\eta, \beta_\eta)$ prior is assumed for $\eta$. In this setup -- this is also done in Lin and Wang (2010) -- $\eta$ is sampled from the full conditional distribution:
		\begin{equation}
		\eta|\bgamma\sim \Gamma\left(\alpha_\eta + K\;,\; \beta_\eta + \sum_{\ell=2}^{K+1}\gamma_\ell\right).\label{sampleeta}
		\end{equation}
		The spline parameters $\bgamma$ are now sampled univariately as in Algorithm \ref{samplegamma}.
		\begin{algorithm}[h]\setstretch{1.35}
			\caption{Algorithm for sampling spline coefficients marginally from their full posteriors}\label{samplegamma}
			\begin{algorithmic}[1]
				\Inputs{Spline coefficient index $\ell$\\ $\bL$, $\bR$, $\bX$, $\bZ$, $\bB$, $\bgamma_{(-\ell)}$, $\bSigma$, $\bbeta$, $\bkappa$, $d$, $v_0, m_0,\eta$;\\}
				\If{$\ell=1$}
				\State Set $\sigma^2:=(v_0+\bb_1^\top\bSigma^{-1}\bb_1)^{-1};$ 
				\vspace{1mm}\State Set $\mu := \sigma^2\left(m_0v_0 + \bb_1^\top\bSigma^{-1}\left(\bZ-\bX\bbeta-\bB_{(-1)}\bgamma_{(-1)}\right)\right);$
				\State Sample $\gamma_1\sim N(\mu,\; \sigma^2);$
				\Else
				\If{$\bb_{\ell}^\top\bSigma^{-1}\bb_{\ell}>0$}
				\State Set $\sigma^2:=\left(\bb_{\ell}^\top\bSigma^{-1}\bb_{\ell}\right)^{-1};$ 
				\State Set $\mu := \sigma^2\left(\bb_\ell^\top\bSigma^{-1}\left(\bZ - \bX\bbeta-\bB_{(-\ell)}\bgamma_{(-\ell)}\right) - \eta\right);$
				\State Set $$\chi := \underset{\left\{(i,j,k)\;:\;R_{ijk},L_{ijk}\;\in\;(0,\infty)\right\} }{\max}\left(\frac{-Z_{ijk}-\sum_{\ell'\notin \{\ell,1\}} \gamma_{\ell'}\left(B_{\ell'-1}^{d,\bkappa}\left(R_{ijk}\right) - B_{\ell'-1}^{d,\bkappa}\left(L_{ijk}\right)\right)}{B_{\ell-1}^{d,\bkappa}\left(R_{ijk}\right) - B_{\ell-1}^{d,\bkappa}\left(L_{ijk}\right)}\right)^+;$$
				\State Sample $\gamma_\ell\sim N(\mu, \sigma^2)\mathbb{I}(\gamma_\ell>\chi);$
				\Else
				\State Sample $\gamma_\ell\sim \text{Exp}(\eta);$
				\EndIf
				\EndIf\\
				{\bf Outputs:}\\
				$\;\;\;\;\gamma_\ell.$
			\end{algorithmic}
		\end{algorithm}
	\end{itemize}
	\FloatBarrier
	
	\section{Gibbs sampler Output Analysis \label{resultssimstudies}}
	This section describes the setup and outcomes of both simulation studies. For both simulation studies and the real data application, all prior parameters except those for $\tau_2$ were set to zero, leading to improper uniform priors for $(\bbeta,\gamma_1)$, the prior $p(\eta) \propto 1/\eta $ for $\eta$ and improper inverse gamma prior $p(\tau_1)\propto 1/(\tau_1+1/n_0)$  for the intra subject covariance. 
		For the first simulation study $\alpha_{\tau_2}=\beta_{\tau_2}=0$ was used, leading to the conditional prior $p(\tau_2|\tau_1)\propto 1/(\tau_2 + \tau_1/\bar{n}_1 + 1/(n_0\bar{n}_1))$, for the second simulation study and real-life data application $\alpha_{\tau_2}=\beta_{\tau_2}=0.001$ was taken.
	
	\subsection{Many Treatment Groups}
	The performance of the Gibbs sampling algorithm (Appendix \ref{Gibbssampler}) was evaluated, to fit the semi-parametric multivariate probit model with a two-way nested covariance structure (Equation (20)), for a large number of treatment groups ($n_2=100$). The number of event types was set to $n_0=5$. An unbalanced design was defined, where group sizes $n_{1i}$ were sampled from a truncated Poisson distribution with mean $5$ ($\min=2$, $\max=\bar{m} = 10$). Five covariates were included and covariate effects $\bm{\beta}$ were sampled uniformly from the interval $[-1,1]$. The first three covariates were sampled from a standard normal distribution, and the last two covariates were sampled from a Bernoulli distribution with success probability $0.5$. Covariates were sampled uniquely on a subject level, and are hence equal for events from the same subject. Interval endpoints were taken with steps $0.1$ on the interval $[0,30].$ Additionally, at random $1\%$ of the measurements were left- or right-censored. The true baseline $h_{ijk}$ was sampled as a linear combination of monotone splines with a degree of four, and the baseline function ranged from $-6$ to $9$. For each value of $\tau_2=-0.2, -0.1,-0.05,\;0,\;0.05,\;0.1,\;0.2,\dots,0.5$ a total of $1,000$ data replications were made. Given $\tau_2$, covariance parameter $\tau_1$ was sampled uniformly from the interval $[\tau_{1L},\tau_{1L}+0.5]$, where 
	\begin{equation}
	\tau_{1L} = -1/n_0 + \max(0,-\max_in_{1i}\tau_2),\label{lbrho}
	\end{equation}
	which ensured that constraints (24) and (25) were satisfied.

		The burn-in period was set to $3,000$ iterations. After that, the MCMC algorithm was halted after both obtaining $6,000$ MCMC draws and an effective sample size of $100$ for all non-spline parameters. The effective sample size was computed using the \verb|R| package \verb|coda| \cite{codapckg} (function \verb|effectiveSize|). For an additional summary of the outcomes of the first simulation study, see Figure \ref{firststudy_summary} below.
	
	\subsubsection*{Parameter Recovery Results}
	In Table \ref{paramrecov1}, the parameter recovery results are shown. For the covariate effects $\bm{\beta}$, the reported 95\% coverage rate (CR, using the 95\% highest posterior density credible interval), (median) coefficient of variation (CV), and (median) relative bias (RB), are the averaged values for all $\bm{\beta}$ parameters. The RB was computed as the bias (estimate - true) relative to the absolute true value of the parameters. The posterior median of the CV and RB were reported. Furthermore, for $\tau_2$, the posterior median and standard deviation were calculated (also computing the median across replications) under the label median and SD, respectively.
	
	The results show that the posterior samples describe the true parameters, under which the data was sampled, well. This is observed from the point estimates for $\tau_2$, which are close to the true values. Furthermore, the RB and CB are quite small and the CR are close to the $95\%$ level (see Table \ref{paramrecov1}). The parameter recovery results slightly decrease in quality when $\tau_2$ is negative. This might be due to sampling relatively high values for $\tau_1$ due to the lower bound in Equation \eqref{lbrho}. The higher covariance might have increased the autocorrelation between latent variables in different Gibbs sampling iterations. Furthermore, the increase of $\tau_1$ could have increased the posterior variance estimate of $\bm{\beta}$.
	\begin{table}[h!]
	\centering
		\caption{Parameter recovery results for many treatment groups: Estimation results for different true values of $\tau_2$.} \label{paramrecov1}
		\begin{tabular}{lccccccccc}
\hline
 & \multicolumn{3}{c}{$\bm{\hat{\beta}}$} & \multicolumn{3}{c}{$\hat{\tau}_1$} & \multicolumn{3}{c}{$\hat{\tau}_2$} \\ 
$\tau_2$  & CP95 & CV & RB & CP95 & CV & RB & CP95 & median & \multicolumn{1}{c}{SD} \\ 
\hline
\nopagebreak -0.2  & $0.928$ & $0.280$ & $\phantom{-}0.007$ & $0.907$ & $0.083$ & $-0.018$ & $0.898$ & $-0.197$ & $0.024$ \\
\nopagebreak -0.1  & $0.932$ & $0.215$ & $\phantom{-}0.001$ & $0.934$ & $0.091$ & $-0.013$ & $0.911$ & $-0.100$ & $0.016$ \\
\nopagebreak -0.05  & $0.934$ & $0.167$ & $\phantom{-}0.001$ & $0.929$ & $0.105$ & $-0.013$ & $0.908$ & $-0.050$ & $0.012$ \\
\rule{0pt}{1.7\normalbaselineskip}0  & $0.948$ & $0.124$ & $\phantom{-}0.004$ & $0.922$ & $0.135$ & $-0.010$ & $0.927$ & $\phantom{-}0.000$ & $0.011$ \\
\nopagebreak 0.05  & $0.947$ & $0.123$ & $\phantom{-}0.001$ & $0.933$ & $0.136$ & $\phantom{-}0.002$ & $0.942$ & $\phantom{-}0.049$ & $0.018$ \\
\nopagebreak 0.1  & $0.946$ & $0.131$ & $-0.002$ & $0.945$ & $0.137$ & $\phantom{-}0.003$ & $0.956$ & $\phantom{-}0.098$ & $0.026$ \\
\rule{0pt}{1.7\normalbaselineskip}0.2  & $0.943$ & $0.129$ & $-0.004$ & $0.945$ & $0.141$ & $\phantom{-}0.010$ & $0.947$ & $\phantom{-}0.197$ & $0.042$ \\
\nopagebreak 0.3  & $0.942$ & $0.132$ & $\phantom{-}0.006$ & $0.941$ & $0.137$ & $\phantom{-}0.000$ & $0.940$ & $\phantom{-}0.296$ & $0.057$ \\
\nopagebreak 0.4  & $0.944$ & $0.138$ & $\phantom{-}0.002$ & $0.933$ & $0.141$ & $\phantom{-}0.004$ & $0.946$ & $\phantom{-}0.395$ & $0.072$ \\
\rule{0pt}{1.7\normalbaselineskip}0.5  & $0.944$ & $0.134$ & $-0.002$ & $0.951$ & $0.139$ & $\phantom{-}0.006$ & $0.941$ & $\phantom{-}0.501$ & $0.087$ \\
\hline 
\end{tabular}

		\noindent $CR$: $95\%$ empirical coverage rate, $CV$: median coefficient of variation, $RB$: median relative bias, $SD$: posterior standard deviation.
	\end{table} \FloatBarrier
	In Table \ref{covgtabl1}, for different levels of highest posterior credibility, the coverage rates are reported for the true values of $\bm{\beta}$, $\tau_1$, and $\tau_2$ under which the data was generated. The coverage rates are averaged across scenarios and the reported coverage for $\bm{\beta}$ is the averaged coverage for all marginal covariate effects. It can be seen that the coverage rates lie close to the specified levels. This shows that the posterior samples accurately describe the posterior distribution of the parameters. 
	\begin{table}[h!]
 		\centering
		\caption{Parameter recovery results for many treatment groups: Coverage rates for $\bm{\beta}$, $\tau_1$, and $\tau_2$ for different credibility levels.}\label{covgtabl1}
		\begin{tabular}{lcccccc}
\hline
param  & 0.6 & 0.7 & 0.8 & 0.9 & 0.95 & \multicolumn{1}{c}{0.99} \\ 
\hline
\nopagebreak $\beta$  & $0.59$ & $0.69$ & $0.79$ & $0.89$ & $0.94$ & $0.98$ \\
\nopagebreak $\tau_1$  & $0.57$ & $0.67$ & $0.77$ & $0.88$ & $0.93$ & $0.98$ \\
\nopagebreak $\tau_2$  & $0.57$ & $0.67$ & $0.77$ & $0.88$ & $0.93$ & $0.98$ \\
\hline 
\end{tabular}

	\end{table}
	
\begin{figure}[h!]
		\centering
		\includegraphics[width = 0.85\textwidth]{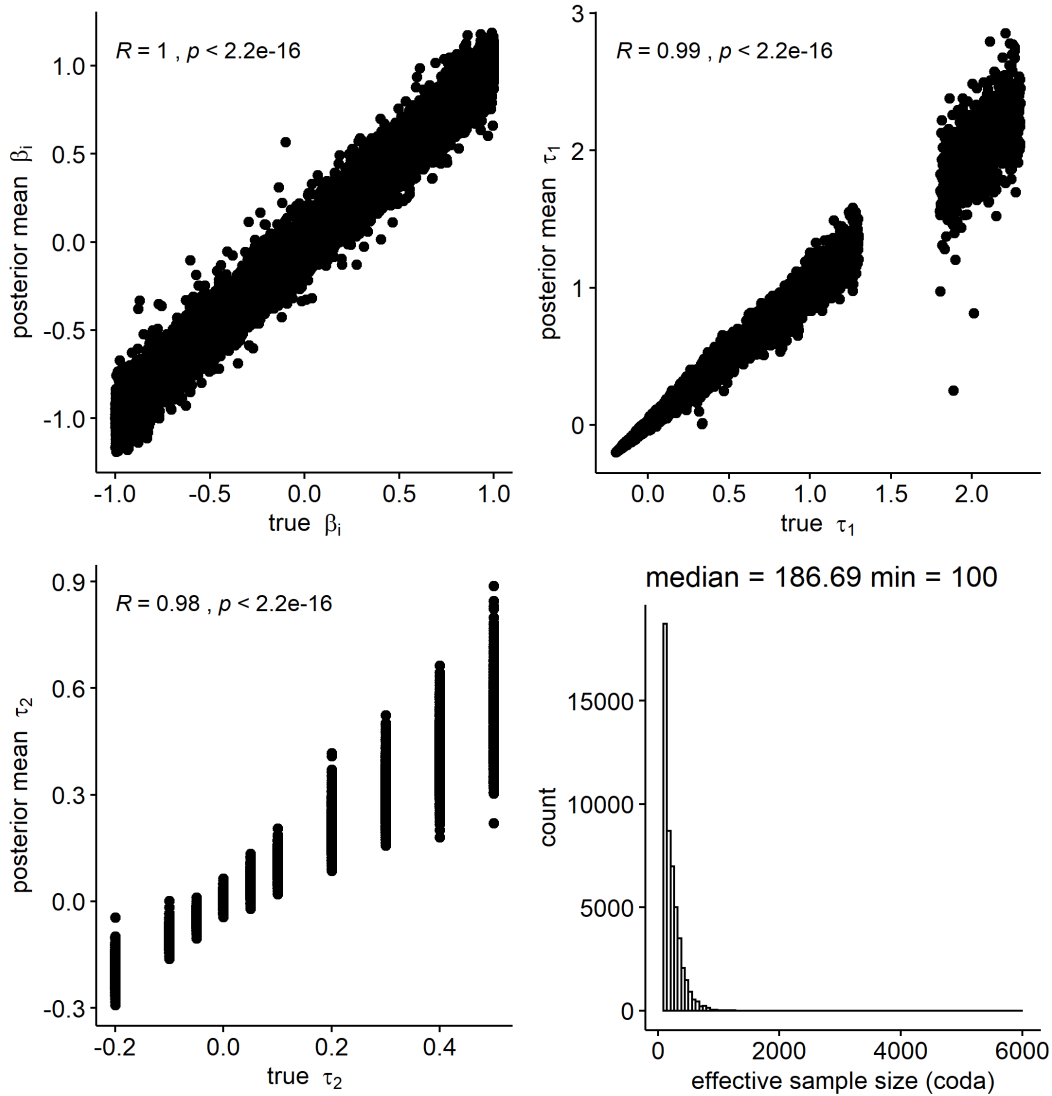}
		\caption{Summary of MCMC results in the first simulation study. $R$ denotes Pearson's correlation coefficient. The excluded values in the support of $\tau_1$ are due to a larger jump in the lower bound of $\tau_1$ in \eqref{lbrho} when going from $\tau_2 = -0.1$ to $\tau_2 = -0.2$. The effective sample sizes are calculated using the function effectiveSize in the R-package coda.}
		\label{firststudy_summary}
	\end{figure}

	 In Figure \ref{firststudy_summary}, Gibbs sampler output results are shown for all simulations in the first simulation study. In all figures except the bottom right one, estimated posterior means are plotted vs. the true parameter values for all scenarios. For the covariate effects $\beta_i$, the results are aggregated in the top left plot. The jump in the support of $\tau_1$ that is seen in the top right figure is due to the large jump in the lower bound \eqref{lbrho} when going from $\tau_2 = -0.1$ to $\tau_2 = -0.2$.  In each plot the Pearson correlation coefficient $R$ is shown, which lies around $1$ for all three datasets. The reported P-values are based on the result stated on e.g. page 10 in Pitman (1939) that a transformation of $R$ follows a Student's t-distribution when assuming no correlation. It is seen that the estimated posterior means are centered around the true parameter value for all three plots.\\
	In the bottom right figure, the distribution of the effective sample sizes is shown for all MCMC outputs in the first study. For any given scenario, the effective sample size is calculated as the minimum univariate effective sample size estimated on the MCMC output for all non-spline parameters $(\bbeta,\tau_1,\tau_2)$. The univariate effective sample size is calculated using the \verb|effectiveSize| function from the \verb|R|-package \verb|coda|. The mean absolute sample size was $6250$ while the median effective sample size was $186.69$.

	\FloatBarrier
	
	\subsection{Few Treatment Groups}
	Conditions similar to our real-data application were considered to evaluate the performance of the Gibbs sampler and the quality of inferences under the model (Equation (20)). Three event types were considered and three treatment groups. The treatment group sizes were $n_{11}=1169$, $n_{12} =1173$, and $n_{13}=1172$. The lower bound (25) is approximately $-6.3\cdot 10^{-4}$ and hence very close to zero. A total of $500$ data replications was made for each value of $\tau_2= \{0, 0.025,0.05,0.075,0.1,0.2,0.3\}.$ The parameter $\tau_1$ was fixed to $0.4$. The number of covariates was five and the covariate effects were sampled uniformly on the interval $[-1,1]$. The first three covariates were sampled from a standard normal distribution and the last two covariates from a Bernoulli distribution with success probability $0.5$. Covariates were sampled uniquely on a subject level, and are hence equal for events from the same subject. The measurement times (support of $\bL,\bR$) were taken with steps $1$ on the interval $[0,365]$. Left- or right-censoring was not applied in this study.
	
\begin{figure}[h]
		\centering
		\includegraphics[width = 0.80\textwidth]{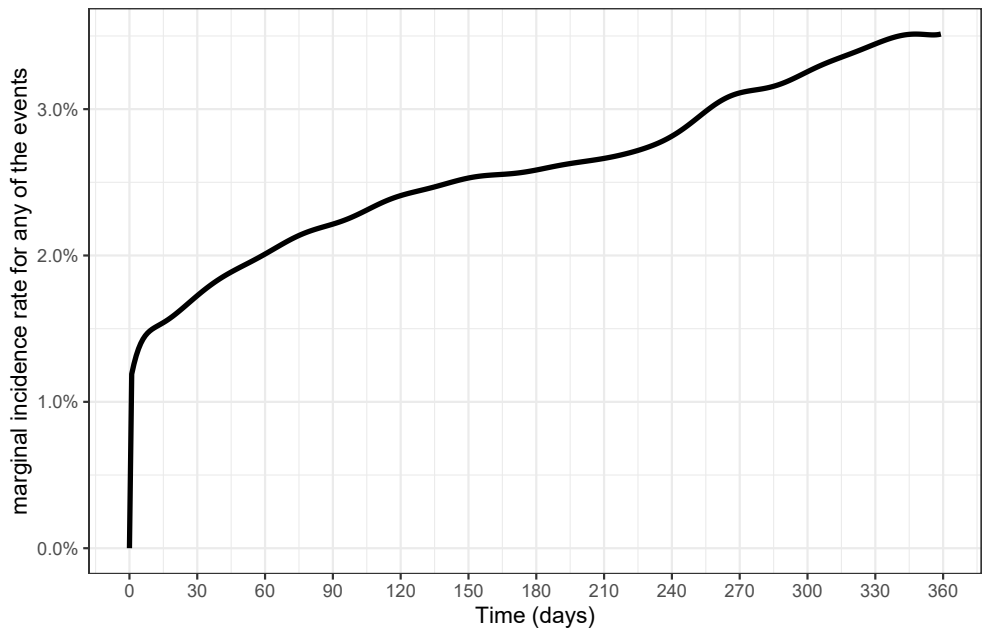}
		\caption{Simulation study for a few treatment groups: Cumulative distribution function of occurrence of an event of any type when $\tau_1=\tau_2=0$.}\label{cdfET}
	\end{figure}

	The baseline function was determined by performing spline regression on an estimate of $h$ from the real data study. The true baseline function was kept fixed across data replications. Under these specifications and for $\bm \tau=\bm 0$, the cumulative distribution of the implied marginal incidence rate (for any event type) is shown in Figure \ref{cdfET}. It can be seen that a challenging aspect of the study was that the marginal incidence rate after one year was very low, and around $3.5\%$. 
	
	The burn-in period was set to $3,000$ MCMC iterations. After that, the Gibbs sampler was ended after both  $6,000$ iterations were collected and an effective sample size of $100$ was obtained for all non-spline parameters. The degree of the splines was set to four, while the spline knots were taken at $20$ equidistant values between the lowest and highest measured event time endpoints. The Gibbs sampling output was generated with the use of improper priors for all parameters except for $\tau_2$, where shape and scale parameters were equal to $0.001$. A proper prior was used for $\tau_2$ to avoid sampling (very) high values for $\tau_2$, which could result in numerical issues. For an additional summary of the outcomes of the second simulation study, see Figure \ref{secondstudy_summary} below.
	
	
	\subsubsection*{Parameter Recovery Results}
	In Table \ref{paramrecov2}, the estimated CR, CV, and RB are given for the covariance and covariate effect parameters, where the average statistic values are reported for all covariate effects. For parameter $\tau_2$, the posterior median and trimmed standard deviation were reported, and they are shown under the label median and SD, respectively. The trimmed posterior standard deviation was computed as the posterior standard deviation of the posterior sample with values removed outside the 99\% empirical credible interval. It can be seen that the  posterior samples describe the true parameters $\bm{\beta}$ and $\tau_1$ quite well, as the RB and CV are quite small and the CR are close to $95\%$ (see Table \ref{paramrecov2}). When comparing the results in Table \ref{paramrecov2} with those in Table \ref{paramrecov1}, it can be seen that the results are similar for the considered parameters. Hence, for the challenging condition with a few and large treatment groups and a (very) low incidence rate, it was still possible to make valid inferences about covariate effects and the intra-subject covariance. 
	
	The posterior distribution for $\tau_2$ contained relatively little information, since there were only three treatment groups. The median point estimator performed well, but the coverage rates differed slightly from the $95\%$ level for small values of $\tau_2$. For a relatively small amount of clusters and small true value for $\tau_2$, the posterior distribution of $\tau_2$ had most of the probability mass in a small interval around zero. This led to small SDs for small $\tau_2$ values. However, for higher $\tau_2$ values, the SD quickly increased. As expected, the SDs for $\tau_2$ in Table \ref{paramrecov2} are much larger than those reported in Table \ref{paramrecov1}.
	
	\begin{table}[h]
		\centering
		\caption{Simulation study for a few treatment groups: Estimation results for different true values of $\tau_2$.}
		\label{paramrecov2}
		\begin{tabular}{lccccccccc}
\hline
 & \multicolumn{3}{c}{$\bm{\hat{\beta}}$} & \multicolumn{3}{c}{$\hat{\tau}_1$} & \multicolumn{3}{c}{$\hat{\tau}_2$} \\ 
$\tau_2$  & CP95 & CV & RB & CP95 & CV & RB & CP95 & median & \multicolumn{1}{c}{SD} \\ 
\hline
\nopagebreak 0  & $0.952$ & $0.121$ & $\phantom{-}0.002$ & $0.936$ & $0.131$ & $0.010$ & $1.000$ & $-0.001$ & $0.005$ \\
\nopagebreak 0.025  & $0.945$ & $0.109$ & $-0.005$ & $0.952$ & $0.130$ & $0.013$ & $0.866$ & $\phantom{-}0.022$ & $0.239$ \\
\nopagebreak 0.05  & $0.955$ & $0.117$ & $-0.001$ & $0.950$ & $0.127$ & $0.014$ & $0.904$ & $\phantom{-}0.048$ & $0.458$ \\
\rule{0pt}{1.7\normalbaselineskip}0.075  & $0.947$ & $0.119$ & $-0.001$ & $0.938$ & $0.130$ & $0.028$ & $0.926$ & $\phantom{-}0.076$ & $0.696$ \\
\nopagebreak 0.1  & $0.946$ & $0.116$ & $-0.001$ & $0.940$ & $0.131$ & $0.020$ & $0.948$ & $\phantom{-}0.106$ & $0.959$ \\
\nopagebreak 0.2  & $0.949$ & $0.119$ & $\phantom{-}0.001$ & $0.948$ & $0.131$ & $0.020$ & $0.944$ & $\phantom{-}0.206$ & $1.825$ \\
\rule{0pt}{1.7\normalbaselineskip}0.3  & $0.939$ & $0.114$ & $\phantom{-}0.004$ & $0.944$ & $0.129$ & $0.022$ & $0.934$ & $\phantom{-}0.295$ & $2.789$ \\
\hline 
\end{tabular}

		\noindent $CR$: $95\%$ empirical coverage rate, $CV$: median coefficient of variation, $RB$: median relative bias, $SD$: trimmed posterior standard deviation.
	\end{table}
	
	In Table \ref{covgtabl2}, for different confidence levels, the coverage rates are shown for the true values of $\bm{\beta}$, $\tau_1$, and $\tau_2$. The coverage rates were averaged over all replications and the reported coverage for $\bm{\beta}$ was the average coverage rate for all covariate effects. It can be seen that the coverage rates lie close to the confidence levels for the parameters $\bm{\beta}$ and $\tau_1$. The posterior samples give a good description of the posterior distribution for these parameters. For small confidence levels, the coverage rates for $\tau_2$ are lower than expected. Most of the discrepancy in coverage rates occurred when $\tau_2$ was close to zero. 
	
	\begin{table}
		\centering
		\caption{Simulation study for a few treatment groups: coverage rates for $\beta$, $\tau_1$, and $\tau_2$ for different confidence levels.}\label{covgtabl2}
		\begin{tabular}{lcccccc}
\hline
param  & 0.6 & 0.7 & 0.8 & 0.9 & 0.95 & \multicolumn{1}{c}{0.99} \\ 
\hline
\nopagebreak $\beta$  & $0.60$ & $0.70$ & $0.80$ & $0.90$ & $0.95$ & $0.99$ \\
\nopagebreak $\tau_1$  & $0.60$ & $0.70$ & $0.80$ & $0.89$ & $0.94$ & $0.99$ \\
\nopagebreak $\tau_2$  & $0.54$ & $0.63$ & $0.75$ & $0.89$ & $0.93$ & $0.97$ \\
\hline 
\end{tabular}

	\end{table}
	
\begin{figure}[h!]
		\centering
		\includegraphics[width = 0.85\textwidth]{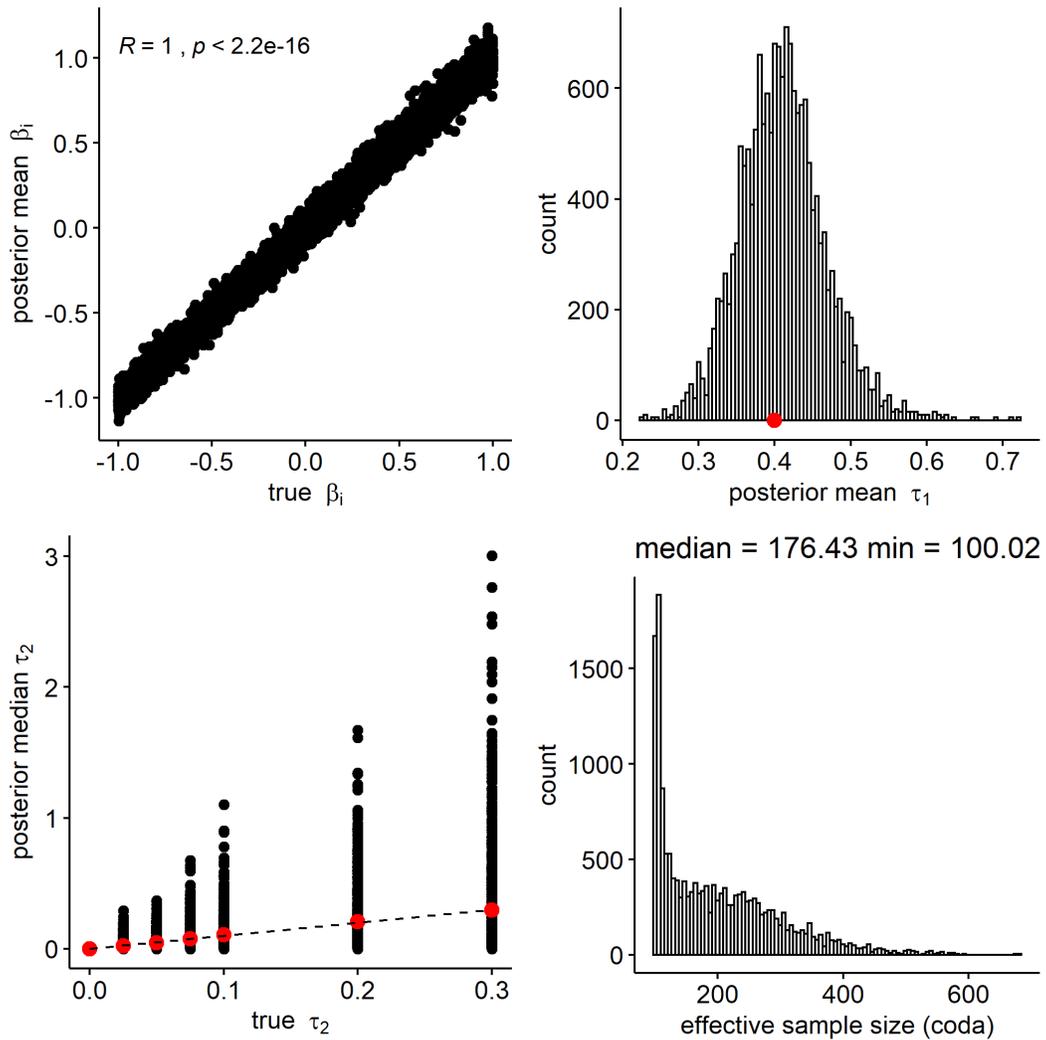}
		\caption{Summary of MCMC results in the second simulation study. $R$ denotes pearson's correlation coefficient, the dots in the top right and bottom left figure denote posterior sample medians in the second and third plot. The dashed line in the third plot denotes a line through the true values of $\tau_2$. The effective sample sizes are calculated using the function effectiveSize in the R-package coda.}
		\label{secondstudy_summary}
	\end{figure}

	In Figure \ref{secondstudy_summary}, Gibbs sampler output results are shown for all simulations in the second simulation study. The top right figure again shows the estimated posterior means of $\beta_i$ vs. their true value. Again, the correlation is close to $1$. The spread around the true value is smaller due to the larger number of subjects enrolled, leading to a larger number of unique covariate values. In the top right figure, a histogram is given for the estimated posterior mean for $\tau_1$. The median of these estimates (red dot) is seen to lie close to 0.4. \\
	In the bottom left figure, the estimated posterior median for $\tau_2$ is shown vs. the true value of $\tau_2$. It is seen that the posterior estimates of $\tau_2$ vary a lot over the scenarios. The median of medians is displayed by red dots, and the dashed line goes through the true values of $\tau_2$. It is hence empirically seen that the estimated posterior means of $\tau_2$ lie below/above the true value with probability approximately $50\%.$ Something that also stands out from this plot is the low spread in the posterior means when $\tau_2=0.$\\
	In the bottom right figure, a histogram is again given of the effective sample sizes. The median absolute sample size was again $6250$ and the median effective sample size was $176.43$. It is seen that the maximum obtained effective sample size is lower in the second simulation study as compared to the first.

\end{document}